\documentclass{ptephy}

\allowdisplaybreaks

\Year{2015}

\begin{document}

\title{Calculation of radiation reaction effect on orbital parameters
in Kerr spacetime}

\author{\name{Norichika Sago}{1,\ast}, and \name{Ryuichi Fujita}{2}}

\address{\affil{1}{Faculty of Arts and Science, Kyushu University,
Fukuoka 819-0395, Japan}
\affil{2}{CENTRA, Departamento de F\'{\i}sica, 
Instituto Superior T\'ecnico, Universidade de Lisboa,
Avenida Rovisco Pais 1, Portugal}
\email{sago@artsci.kyushu-u.ac.jp}}

\begin{abstract}
We calculate the secular changes of the orbital parameters of a point particle
orbiting a Kerr black hole, due to the gravitational radiation reaction.
For this purpose, we use the post-Newtonian (PN) approximation in the first
order black hole perturbation theory, with the expansion with respect to the
orbital eccentricity.
In this work, the calculation is done up to the fourth post-Newtonian (4PN)
order and to the sixth order of the eccentricity, including the effect of the
absorption of gravitational waves by the black hole.
We confirm that, in the Kerr case, the effect of the absorption
appears at the 2.5PN order beyond the leading order in the secular change
of the particle's energy and may induce a superradiance, as known previously
for circular orbits.
In addition, we find that the superradiance may be suppressed when the orbital
plane inclines with respect to the equatorial plane of the central black hole.
We also investigate the accuracy of the 4PN formulae
by comparing to numerical results. 
If we require that the relative errors in the 4PN formulae are less than
$10^{-5}$, 
the parameter region to satisfy the condition will be $p\gtrsim 50$ for $e=0.1$, 
$p\gtrsim 80$ for $e=0.4$, and $p\gtrsim 120$ for $e=0.7$
almost irrespective of the inclination angle nor the spin of the black hole,
where $p$ and $e$ are the semi-latus rectum and the eccentricity of the orbit.
The region can further be extended using an exponential resummation method
to $p\gtrsim 40$ for $e=0.1$, $p\gtrsim 60$ for $e=0.4$, 
and $p\gtrsim 100$ for $e=0.7$.
Although we still need the higher order calculations of the PN approximation
and the expansion with respect to the orbital eccentricity 
to apply for data analysis of gravitational waves, 
the results in this paper would be an important improvement from 
the previous work at the 2.5PN order, especially for large $p$ region.
\end{abstract}

\subjectindex{E01, E02, E31, E36} 

\maketitle

\section{Introduction}
The gravitational two-body problem is a fundamental issue
in general relativity. This also attracts great interest in
gravitational wave physics because binary inspirals are promising
sources of gravitational waves which are expected to be detected
directly by ongoing gravitational wave observatories in the world.
Understanding the dynamics of binary system is required to
predict the emitted gravitational waveforms accurately for
efficient searches of the signal in observed data.

One of major approaches for this purpose is the gravitational
self-force (GSF) picture in the black hole perturbation theory.
In this picture, a binary is regarded as a point mass orbiting
a black hole and the dynamics can be described by the equation
of motion of the mass including the effect of the interaction
with the self-field, that is, the GSF.
After the formal expression of the GSF was presented by
Mino, Sasaki and Tanaka \cite{Mino:1996nk} and Quinn and Wald
\cite{Quinn:1996am},
a lot of efforts have been devoted to develop practical formulations
and methods to calculate the GSF
(for example, refer to \cite{Poisson:2011nh} for the formulation of
GSF, \cite{Barack:2009ux, Wardell:2015kea}
for the recent progress in practical calculations of GSF).

Although a lot of progress has been made, however, it is still
challenging to calculate the GSF directly  for general orbits,
especially in Kerr spacetime.
Practical calculations of the GSF with high accuracy will require a
huge amount of time and computer resources mainly because of the
regularization problem induced by the point mass limit.
Therefore it is important to develop a way to reduce the cost of
computing the GSF.
The two-timescale expansion method \cite{Hinderer:2008dm} gives
a hint for it:
assuming that a point mass does not encounter any transient resonances
(\textit{e.g.} shown in \cite{Flanagan:2010cd}),
the orbital phase, which is the most important information to predict
the waveform, can be expressed in the expansion with respect to the
mass ratio, $\eta$, as
\begin{equation}
\Phi =
 \eta^{-1} \left[ \Phi^{(0)} + \eta \Phi^{(1)} + O(\eta^2) \right],
\end{equation}
where $\Phi^{(0)}$ and $\Phi^{(1)}$ are quantities of order unity.
The leading term, $\Phi^{(0)}$, can be calculated from the
knowledge up to the time-averaged dissipative piece of the first order
GSF, corresponding to the secular growth.
The calculation of this secular contribution can be simplified
significantly by using the radiative field defined as half the
retarded solution minus half the advanced solution for the
equation of the gravitational perturbation
\cite{Mino:2003yg, Sago:2005gd, Sago:2005fn}, {\it i.e.}
the adiabatic approximation method, because the radiative field
is the homogeneous solution free from the divergence induced by
the point mass limit. This method allows us to
calculate the leading term  accurately without spending huge
computational resources.
On the other hand, the calculation of $\Phi^{(1)}$ requires the rest of
the first order GSF (the oscillatory part of the dissipative
GSF and the conservative GSF) and the time-averaged dissipative piece
of the second order GSF. There is no simplification in calculating
these post-1 adiabatic pieces at present. Since $\Phi^{(1)}$ is
subleading, however, the requirement of the accuracy is not so high
compared to that of the leading term.
This fact suggests that it is possible to reduce the computational
cost by using a suitable method with an appropriate error tolerance
to calculate each piece of the GSF (for example, a hybrid approach
is proposed in \cite{Osburn:2014hoa}).

In this work, we focus on the time-averaged dissipative part of the
first order GSF, which 
has the dominant contribution to the evolution of inspirals,
and present the analytic post-Newtonian (PN) formulae.
So far, several works in this direction had been done for two
restricted classes of orbits: circular orbits and equatorial orbits.
(See \cite{Mino:1997bx} and references therein for early works
in 1990's).
Recently, thanks to the progress of computer technology,
the very higher order post-Newtonian calculations can be possible
for circular equatorial orbits:
the 22PN calculation of the energy flux is demonstrated in Schwarzschild
case \cite{Fujita:2012cm}, and the 11PN calculation in Kerr case
\cite{Fujita:2014eta}.
There is also the calculation of the secular GSF effects for slightly
eccentric and slightly inclined (non-equatorial) orbits
\cite{Sago:2005fn}, and later it had been extended to orbits with
arbitrary inclination \cite{Ganz:2007rf}, where
the PN formulae of the secular GSF effects are
presented in the expansion with respect to the orbital eccentricity.
However, the calculation in \cite{Ganz:2007rf} has been done only up
to the 2.5PN order with the second order correction of the eccentricity.
Also the absorption to the black hole is ignored there.
The main purpose of this work is to update the results in
\cite{Ganz:2007rf} up to the 4PN order and the sixth order correction
of the eccentricity, including the effect of the absorption
to the black hole.

This paper is organized as follows.
In Sec.\ref{sec:review}, we give a brief review of the geodesic
motion of a point particle in Kerr spacetime, the gravitational
perturbations induced by the particle, and the adiabatic
approximation method of calculating the secular effect of the GSF.
In Sec.\ref{sec:dotE_formulae}, we present the PN formulae of the
secular changes of the the energy, azimuthal angular momentum,
and Carter parameter of the particle due to the gravitational radiation
reaction in the expansion with respect to the orbital eccentricity.
In Sec.\ref{sec:compare}, we investigate the accuracy of our PN formulae
by comparing to numerical results given by the method in
\cite{Fujita:2004rb,Fujita:2009uz,Fujita:2009us}, which can give each
modal flux at the accuracy about 14 significant figures.
In Sec.\ref{sec:exp-resum}, we implement a resummation
method to the PN formulae given in this work in order to improve the
accuracy.
In Sec.\ref{sec:Delta_n}, we discuss the convergence of the analytic
formulae as the PN expansion and the expansion with respect to
the eccentricity.
Finally, we summarize the paper in Sec.\ref{sec:summary}.
For the readability of the main text,
we present the PN formulae for the orbital parameters, the fundamental
frequencies, the orbital motion in Appendices \ref{App:variable_exp}
and \ref{App:coeff_orbit}, which are used in
calculating the secular changes of the orbital parameters.
We also present the PN formulae for the secular changes of an
alternative set of the orbital parameters
in Appendix \ref{App:orbit_evolv}.
Throughout this paper we use metric signature $(-+++)$ and
``geometrized'' units with $c=G=1$.

\section{Review of formulation: Adiabatic radiation reaction}
\label{sec:review}
The orbital evolution of a point particle due to the time-averaged
dissipative part of the GSF is often described in terms of the secular
changes of the orbital parameters. In order to calculate the changes,
we need the information on the first order gravitational perturbations
induced by the particle when it moves along the background geodesics.
In this section, we review the geodesic dynamics of a point particle
in Kerr spacetime, the gravitational perturbations induced by the
particle, and the adiabatic evolution of the orbital parameters.

\subsection{Geodesic motion}
The Kerr metric in the Boyer-Lindquist coordinates,
$(t, r, \theta, \varphi)$, is given by
\begin{eqnarray}
g_{\mu\nu}dx^\mu dx^\nu &=&
-\left(1-\frac{2Mr}{\Sigma}\right)dt^2
-\frac{4Mar\sin^2\theta}{\Sigma}dtd\varphi
+\frac{\Sigma}{\Delta}dr^2
\nonumber \\ &&
\hspace*{2cm} +\Sigma d\theta^2
+\left(r^2+a^2+\frac{2Ma^2r}{\Sigma}\sin^2\theta\right)
\sin^2\theta d\varphi^2, \label{eq:Kerr}
\end{eqnarray}
where $\Sigma=r^2+a^2\cos^2\theta$, $\Delta=r^2-2Mr+a^2$,
$M$ and $aM$ are the mass and angular momentum of
the black hole, respectively.

There are two Killing vectors related to the stationarity and
axisymmetry of Kerr spacetime, which are expressed as
$\xi_{(t)}^{\mu}=(1,0,0,0)$ and $\xi_{(\varphi)}^{\mu}=(0,0,0,1)$.
In addition, it is known that Kerr spacetime possesses a Killing tensor,
$K_{\mu\nu}=2\Sigma l_{(\mu}n_{\nu)}+r^2g_{\mu\nu}$, where
$l^{\mu}$ and $n^{\mu}$ are the Kinnersley's null vectors
given by
\begin{equation}
l^{\mu} :=
\left(\frac{r^2+a^2}{\Delta},1,0,\frac{a}{\Delta} \right), \quad
n^{\mu}:=
\left(\frac{r^2+a^2}{2\Sigma},-\frac{\Delta}{2\Sigma},
0,\frac{a}{2\Sigma}\right).
\end{equation}
For the geodesic motion of a particle in Kerr geometry, there are
three constants of motion related to the symmetries:
\begin{equation}
\hat{E} :=
-u^{\alpha}\xi_{\alpha}^{(t)}, \quad
\hat{L} :=
u^{\alpha}\xi_{\alpha}^{(\varphi)}, \quad
\hat{Q} :=
K_{\alpha\beta}u^{\alpha}u^{\beta}, \label{eq:COMs_specific}
\end{equation}
where $u^{\alpha}$ is the four velocity of the particle.
$\hat{E}$ and $\hat{L}$ correspond to the specific energy and azimuthal
angular momentum of the particle respectively. $\hat{Q}$ is called as
the Carter constant, which corresponds to the square of the specific
total angular momentum in Schwarzschild case.
These specific variables are measured in units of the particle's mass,
$\mu$. One can recover the expressions in the standard units as
\begin{equation}
E:= \mu\hat{E}, \quad
L:= \mu\hat{L}, \quad
Q:= \mu^2\hat{Q}. \label{eq:COMs}
\end{equation}
There is another definition of the Carter constant, $C\equiv Q-(aE-L)^2$,
which vanishes for equatorial orbits. In this paper, we use $C$ as one of
the orbital parameters, instead of $Q$.

By using these constants of motion, the geodesic equations
can be expressed in the following form as
\begin{eqnarray}
\left(\frac{dr}{d\lambda}\right)^2 =
R(r), \quad
\left(\frac{d\cos\theta}{d\lambda}\right)^2 =
\Theta(\cos\theta), \label{eq:eom_r_theta} \\
\frac{dt}{d\lambda} =
V_{tr}(r) + V_{t\theta}(\theta), \quad
\frac{d\varphi}{d\lambda} =
V_{\varphi r}(r) + V_{\varphi\theta}(\theta), \label{eq:eom_t_phi}
\end{eqnarray}
where we introduced a new parameter $\lambda$ through the relation
$d\lambda=d\tau/\Sigma$, and some functions as
\begin{eqnarray}
P(r)&:=&\hat{E}(r^2+a^2)-a\hat{L}, \\
R(r)&:=&[P(r)]^2-\Delta[r^2+(a\hat{E}-\hat{L})^2+\hat{C}], \\
\Theta(\cos\theta)&:=&
\hat{C} - (\hat{C}+a^2(1-\hat{E}^2)+\hat{L}^2)\cos^2\theta
+ a^2(1-\hat{E}^2)\cos^4\theta, \\
V_{tr}(r) &:=& \frac{r^2+a^2}{\Delta}P(r), \quad
V_{t\theta}(\theta) := -a(a\hat{E}\sin^2\theta-\hat{L}), \\
V_{\varphi r}(r) &:=& \frac{a}{\Delta}P(r), \quad
V_{\varphi\theta}(\theta) :=
-\left(a\hat{E}-\frac{\hat{L}}{\sin^2\theta}\right).
\end{eqnarray}

A generic geodesic orbit in Kerr spacetime can be characterized
by three parameters, $\{E, L, C\}$
\footnote{Strictly speaking, the orbit is also characterized by
the initial position of the particle.
However, the time-averaged
dissipative part of the first order GSF does not affect the initial
position (the other parts of the first order GSF and the higher order
GSF will do) \cite{Hinderer:2008dm}.
Also the secular changes of $\{E,L,C\}$
does not depend on the initial position.
Hence we do not need the information on
the initial position to describe the secular evolution of the
orbit at the order considered in this paper.}.
In the case of a bound orbit,
we can use an alternative set of parameters,
$\{r_p, r_a, \theta_{\rm min}\}$, instead of $\{E, L, C\}$,
where $r_p$ and $r_a$ are the values of $r$ at the periapsis and
apoapsis and $\theta_{\rm min}$ is the minimal value of $\theta$,
respectively. Using this set of parameters, we can describe
the range in which the motion takes place as $r_p \le r \le r_a$
and $\theta_{\rm min} \le \theta \le \pi-\theta_{\rm min}$.
There is another useful choice of parameters used in
\cite{Drasco:2003ky}, $\{p, e, \iota\}$,
defined by
\begin{equation}
p := \frac{2r_p r_a}{M(r_a+r_p)}, \quad
e := \frac{r_a-r_p}{r_a+r_p}, \quad
\cos\iota := \frac{L}{\sqrt{L^2+C}}.
\end{equation}
By analogy to the parametrization used in celestial mechanics,
$p$, $e$, $\iota$ are referred as semi-latus rectum,
orbital eccentricity, orbital inclination angle, respectively.
For later convenience, we also introduce $Y=\cos\iota$ and
$v=\sqrt{1/p}$. Since $v$ corresponds to the magnitude of the
orbital velocity, it can be used as the post-Newtonian parameter.
For example, we call the $O(v^8)$-correction from the leading order
as the fourth order post-Newtonian (4PN) correction.

It is worth noting that, by introducing $\lambda$,
the radial and longitudinal equations of motion in
Eq.(\ref{eq:eom_r_theta}), are
completely decoupled. For an bound orbit, therefore, the radial
and longitudinal motions are periodic with the periods,
$\{\Lambda_r, \Lambda_{\theta}\}$, defined by
\begin{equation}
\Lambda_r = 2\int_{r_p}^{r_a} \frac{dr}{\sqrt{R(r)}}, \quad
\Lambda_{\theta} = 4\int_{\theta_{\rm min}}^{\pi/2}
\frac{d\theta}{\sqrt{\Theta(\theta)}}.
\end{equation}
This means that these motions can be expressed in terms of
Fourier series as
\begin{eqnarray}
r(\lambda) &=&
 p\sum_{n_r=0}^{\infty} \alpha_{n_r} \cos n_r\Omega_r\lambda,
\label{eq:r-fourier} \\
\cos\theta(\lambda) &=&
 \sqrt{1-Y^2} \sum_{n_\theta=0}^{\infty}
\beta_{n_\theta} \sin n_\theta \Omega_\theta \lambda,
\label{eq:theta-fourier}
\end{eqnarray}
where $\Omega_r$ and $\Omega_\theta$ are the radial and
longitudinal frequencies given by
\begin{equation}
\Omega_r := \frac{2\pi}{\Lambda_r}, \quad
\Omega_{\theta} := \frac{2\pi}{\Lambda_{\theta}},
\end{equation}
and we choose the initial values so that
$r(\lambda=0)=r_p$ and $\theta(\lambda=0)=\pi/2$.
\footnote{If the ratio of the radial and longitudinal frequencies
is irrational, we can adjust the origin of $\lambda$
approximately so that the radial and longitudinal oscillations
reach the minima simultaneously at $\lambda=0$ \cite{Mino:2003yg}.
On the other hand, it is not the case
if the ratio is rational, {\it i.e.} the resonance
case. This implies that the secular evolution of a resonant
orbit cannot be described only by the PN formulae derived in
this work \cite{Isoyama:2013yor}.}

Since the temporal and azimuthal equations of motion in
Eq.(\ref{eq:eom_t_phi}) are
divided into the $r$- and $\theta$-dependent parts, the
solutions can be divided into three parts:
the linear term with respect to $\lambda$,
the oscillatory part with period of $\Lambda_r$,
and the oscillatory part with period of $\Lambda_\theta$.
They can be expressed as
\begin{eqnarray}
t(\lambda) &=&  \Omega_t \lambda
+ t^{(r)}(\lambda) + t^{(\theta)}(\lambda); \quad
t^{(A)}(\lambda) := \sum_{n_A=1}^{\infty} \tilde{t}_{n_A}^{(A)}
\sin n_A \Omega_A \lambda,
\label{eq:t-fourier} \\
\varphi(\lambda) &=& \Omega_\varphi \lambda
+ \varphi^{(r)}(\lambda) + \varphi^{(\theta)}(\lambda); \quad
\varphi^{(A)}(\lambda) := \sum_{n_A=1}^{\infty} \tilde{\varphi}_{n_A}^{(A)}
\sin n_A \Omega_A \lambda,
\label{eq:phi-fourier}
\end{eqnarray}
where the index $A$ runs over $\{r, \theta\}$, and
\begin{equation}
\Omega_t :=
\left\langle \frac{dt}{d\lambda} \right\rangle_\lambda, \quad
\Omega_\varphi :=
\left\langle \frac{d\varphi}{d\lambda} \right\rangle_\lambda
\end{equation}
with $\langle \cdots \rangle_\lambda \equiv \lim_{T\rightarrow\infty}
(2T)^{-1}\int_{-T}^T d\lambda\cdots$, representing the time average
along the geodesic. We choose the initial conditions as
$t(\lambda=0)=\varphi(\lambda=0)=0$.
$\Omega_\varphi$ corresponds to the frequency of
the orbital rotation.

In Appendices \ref{App:variable_exp} and \ref{App:coeff_orbit},
we present the PN formulae of
the orbital parameters, $\{E,L,C\}$, the fundamental frequencies,
$\{\Omega_t, \Omega_r, \Omega_\theta, \Omega_\phi\}$, and
the Fourier coefficients of the motions in Eqs.~(\ref{eq:r-fourier}),
(\ref{eq:theta-fourier}), (\ref{eq:t-fourier})
and (\ref{eq:phi-fourier}).

\subsection{Secular evolution of orbital parameters}
The gravitational perturbations in Kerr spacetime can be described
by the Weyl scalar,
$\Psi_{4}$, which satisfies the Teukolsky equation \cite{Teukolsky:1973ha}.
To solve the Teukolsky equation, the method of separation of
variables is often used, in which $\Psi_4$ is decomposed
in the form as
\begin{eqnarray}
 \Psi_4 = \sum_{\ell m} \int d\omega R_{\Lambda}(r)
S_{\Lambda}(\theta){\rm e}^{im\varphi-i\omega t},
\end{eqnarray}
where $S_\Lambda(\theta)$ is the spin-2 spheroidal harmonics
and $\Lambda$ represents a set of indices in the Fourier-harmonic
expansion, $\{\ell,m,\omega\}$.
The separated equation for the radial function is given by
\begin{eqnarray}
&& \left[\Delta^{2}\frac{d}{dr}\left(\Delta^{-1}\frac{d}{dr}\right)
+\left(\frac{K^2+4i(r-M)K}{\Delta}-8i\omega r-\bar{\lambda} \right)\right]
R_{\Lambda}(r) = T_{\Lambda},
\label{eq:radial-Teukolsky}
\end{eqnarray}
where 
$$
K\equiv (r^2+a^2)\omega - ma,
$$ 
$T_\Lambda$ is the source term constructed from the
energy-momentum tensor of the point particle,
and $\bar{\lambda}$ is the eigenvalue determined by the equation 
for $S_{\Lambda}$ (To find the basic formulae for the Teukolsky
formalism used in this paper, refer to the section 2 in
\cite{Sasaki:2003xr} for example).

The amplitudes of the partial waves at the horizon and at infinity
are defined by the asymptotic forms of the solution of the radial
equation as
\begin{eqnarray}
R_{\Lambda}(r\to r_+)
\equiv \mu Z^{{\rm H}}_{\Lambda}\Delta^2 {\rm e}^{-ikr^*}, \quad
R_{\Lambda}(r\to\infty)
\equiv \mu Z^{\infty}_{\Lambda} r^3 {\rm e}^{i\omega r^*}
\label{eq:partial-wave}
\end{eqnarray}
with $r_+\equiv M + \sqrt{M^2-a^2}$ and
$k=\omega-ma/(2Mr_+)$.
Since the spectrum with respect to $\omega$ gets discrete in the
case of a bound orbit, $Z_\Lambda^{{\rm H},\infty}$ take the form
\begin{equation}
Z_\Lambda^{{\rm H}, \infty} =
2\pi\delta(\omega-\omega_{m n_r n_\theta})
\tilde{Z}_{\tilde{\Lambda}}^{{\rm H}, \infty},
\end{equation}
where $\tilde{\Lambda}$ denotes the set of indices,
$\{\ell,m,n_r,n_\theta\}$, and
\begin{equation}
\omega_{m n_r n_\theta} \equiv
\Omega_t^{-1}
\left(
m \Omega_\varphi + n_r \Omega_r + n_\theta \Omega_\theta
\right).
\end{equation}
With these amplitudes, the secular changes of the orbital parameters,
$\{E, L, C\}$, can be expressed by
\begin{eqnarray}
\left\langle \frac{dE}{dt} \right\rangle_t
&=&
- \mu^2 \sum_{\tilde\Lambda}
\frac{1}{4\pi\omega_{m n_r n_\theta}^2}
\left(
\left| \tilde{Z}_{\tilde\Lambda}^\infty \right|^2
+ \alpha_{\ell m}(\omega_{m n_r n_\theta})
\left| \tilde{Z}_{\tilde\Lambda}^{\rm H} \right|^2
\right),
\label{eq:Edot} \\
\left\langle \frac{dL}{dt} \right\rangle_t
&=&
- \mu^2 \sum_{\tilde\Lambda}
\frac{m}{4\pi\omega_{m n_r n_\theta}^3}
\left(
\left| \tilde{Z}_{\tilde\Lambda}^\infty \right|^2
+ \alpha_{\ell m}(\omega_{m n_r n_\theta})
\left| \tilde{Z}_{\tilde\Lambda}^{\rm H} \right|^2
\right),
\label{eq:Ldot} \\
\left\langle \frac{dC}{dt} \right\rangle_t
&=&
- 2\left\langle a^2 E \cos^2\theta \right\rangle_\lambda
\left\langle \frac{dE}{dt} \right\rangle_t
+ 2\left\langle L \cot^2\theta \right\rangle_\lambda
\left\langle \frac{dL}{dt} \right\rangle_t
\nonumber \\ &&
- \mu^3 \sum_{\tilde\Lambda}
\frac{n_\theta \Omega_\theta}
{2\pi\omega_{m n_r n_\theta}^3}
\left(
\left| \tilde{Z}_{\tilde\Lambda}^\infty \right|^2
+ \alpha_{\ell m}(\omega_{m n_r n_\theta})
\left| \tilde{Z}_{\tilde\Lambda}^{\rm H} \right|^2
\right),
\label{eq:Cdot}
\end{eqnarray}
where
\begin{equation}
\alpha_{\ell m}(\omega) =
\frac{256(2Mr_+)^5 k (k^2+4\tilde\epsilon^2)
(k^2+16\tilde\epsilon^2) \omega^3}
{|{\cal C}_{S}|^2}, \quad
\tilde\epsilon=\sqrt{M^2-a^2}/(4Mr_+),
\end{equation}
and ${\cal C}_{S}$ is the Starobinsky constant given
by~\cite{Teukolsky:1974yv}
\begin{eqnarray}
|{\cal C}_{S}|^2 &=&
\left[ (\bar{\lambda}+2)^2 + 4a\omega m - 4a^2\omega^2 \right]
\left[ \bar{\lambda}^2 + 36a\omega m -36 a^2\omega^2 \right]
\nonumber \\ &&
+ (2\bar{\lambda}+3)(96a^2\omega^2 - 48a\omega m)
+ 144\omega^2(M^2-a^2).
\end{eqnarray}
It should be noted that, in these formulae, the averaged rates of
change are expressed with respect to the Boyer-Lindquist time,
which can be related to those with respect to $\lambda$
\cite{Drasco:2005is} as
\begin{equation}
\left\langle \frac{dI}{dt} \right\rangle_t =
\left\langle \frac{dt}{d\lambda} \right\rangle_\lambda^{-1} 
\left\langle \frac{dI}{d\lambda} \right\rangle_\lambda
\end{equation}
for a function of time, $I(t)$. Also it should be noted that
each formula in Eqs.~(\ref{eq:Edot})-(\ref{eq:Cdot}) can be divided
into the infinity part and the horizon part: the former consists
of the terms including the amplitudes of the partial waves at the
infinity, $\tilde{Z}_{\tilde{\Lambda}}^{\infty}$, the latter consists
of the terms including the amplitudes at the horizon,
$\tilde{Z}_{\tilde{\Lambda}}^{{\rm H}}$.
As for the energy and azimuthal angular momentum, the infinity
parts are balanced with the corresponding fluxes radiated to infinity
and the horizon parts with the absorption of the gravitational
waves into the central black hole \cite{Teukolsky:1974yv, Gal'tsov82}.

The practical calculation of $\tilde{Z}_{\tilde{\Lambda}}^{{\rm H},\infty}$
involves solving the geodesic equations,
calculating two independent homogeneous solutions of
Eq.(\ref{eq:radial-Teukolsky}) and the spin-2 spheroidal harmonics,
and the Fourier transformation of functions consisting of them.
In this work, we followed the same procedure proposed in
\cite{Ganz:2007rf} to perform these calculations analytically.

In performing the summation in Eqs.~(\ref{eq:Edot})-(\ref{eq:Cdot})
practically, we need to truncate the summation to finite ranges of
$\tilde\Lambda=\{\ell,m,n_r,n_\theta\}$. To obtain the accuracy
of the 4PN and $O(e^6)$, it is necessary to sum $\ell$ in
the range $2\le\ell\le 6$ ($2\le\ell\le 3$), 
$n_r$ in the range $-3\le n_r\le 3$ ($-2\le n_r\le 3$) and 
$n_\theta$ in the range $-8\le n_\theta\le 12$ ($-4\le n_\theta\le 6$) 
for the infinity (horizon) part. The other modes out of these
ranges are the higher PN corrections than the 4PN order or
the higher order corrections than $O(e^6)$.

\section{Results} \label{sec:result}

\subsection{PN formulae of the secular changes of orbital parameters}
\label{sec:dotE_formulae}
In this work, we derived the analytic 4PN order formulae of
Eqs.(\ref{eq:Edot})-(\ref{eq:Cdot}) in the expansion with respect
to the orbital eccentricity, $e$, up to $O(e^6)$
(we simply call them as the 4PN $O(e^6)$ formulae).
Since the full expressions of the 4PN $O(e^6)$ formulae are too
lengthy to show in the text, we show the infinity parts
up to the 3PN order and the horizon parts up to the
3.5PN order (while we keep the expansions with respect to $e$ up
to $O(e^6)$).
The complete expressions of the 4PN $O(e^6)$ formulae 
will be publicly available online~\cite{BHPC}.

The infinity parts of Eqs.(\ref{eq:Edot})-(\ref{eq:Cdot})
are given by
\begin{eqnarray}
\left\langle \frac{dE}{dt} \right\rangle_t^\infty &=&
\left( \frac{dE}{dt} \right)_{\rm N}
\biggl[ 1+{\frac {73}{24}}\,{e}^{2}+{\frac {37}{96}}\,{e}^{4}+ \left\{ -{\frac 
{1247}{336}}-{\frac {9181}{672}}\,{e}^{2}+{\frac {809}{128}}\,{e}^{4}+
{\frac {8609}{5376}}\,{e}^{6} \right\} {v}^{2}
\nonumber\\ \nonumber && 
+ \biggl\{ 4\,\pi-{\frac {73}{12}}\,Yq + 
\left( {\frac {1375}{48}}\,\pi -{\frac {823}{24}}\,Yq \right) {e}^{2}
\\ \nonumber && \hspace{0.5cm}
+ \left( {\frac {3935}{192}}\,\pi -{\frac {949}{32}}\,
Yq \right) {e}^{4}+ \left( {\frac {10007}{9216}}\,\pi -{\frac {491}{
192}}\,Yq \right) {e}^{6} \biggr\} {v}^{3}
\\ \nonumber && 
+ \biggl\{ -{\frac {44711}{9072}}+{\frac {527}{96}}\,{Y}^{2}{q}^{2}-{\frac {329}{96}}\,{q}^{2}+
 \left( -{\frac {172157}{2592}}-{\frac {4379}{192}}\,{q}^{2}+{\frac {
6533}{192}}\,{Y}^{2}{q}^{2} \right) {e}^{2}
\\ \nonumber && \hspace{0.5cm}
+ \left( -{\frac {2764345}{
24192}}-{\frac {3823}{256}}\,{q}^{2}+{\frac {6753}{256}}\,{Y}^{2}{q}^{
2} \right) {e}^{4}
\\ \nonumber && \hspace{0.5cm}
+ \left( {\frac {3743}{2304}}-{\frac {363}{512}}\,{q
}^{2}+{\frac {2855}{1536}}\,{Y}^{2}{q}^{2} \right) {e}^{6} \biggr\} {v}^{4}
\\ \nonumber && 
+ \biggl\{ {\frac {3749}{336}}\,Yq-{\frac {8191}{672}}\,\pi +
 \left( -{\frac {44531}{336}}\,\pi +{\frac {1759}{56}}\,Yq \right) {e}^{2}
\\ \nonumber && \hspace{0.5cm}
- \left( {\frac {4311389}{43008}}\,\pi +{\frac {111203}{1344}}\,Y
q \right) {e}^{4}+ \left( {\frac {15670391}{387072}}\,\pi -{\frac {
49685}{448}}\,Yq \right) {e}^{6} \biggr\} {v}^{5}
\\ \nonumber && 
+ \biggl\{ {\frac {6643739519}{69854400}}-{\frac {1712}{105}}\,\gamma
-{\frac {3424}{105}}\,\ln  \left( 2 \right)+\frac{16}{3}\,\pi^{2} 
+{\frac {135}{8}}\,{q}^{2}-{\frac {169}{6}}\,\pi \,Yq
\\ \nonumber && \hspace{0.5cm}
+{\frac {73}{21}}\,{Y}^{2}{q}^{2}
+ \biggl( {\frac {43072561991}{27941760}}
+{\frac {680}{9}}\,{\pi }^{2}-{\frac {234009}{560}}\,\ln  \left( 3 \right) 
-{\frac {14552}{63}}\,\gamma
\\ \nonumber && \hspace{1.0cm}
-{\frac {13696}{315}}\,\ln  \left( 2 \right)
+{\frac {205747}{1344}}\,{q}^{2}-{\frac {4339}{16}}\,\pi \,Yq+{\frac {13697}{192}}\,{Y}^{2}{q}^{2} \biggr) {e}^{2}
\\ \nonumber && \hspace{0.5cm}
+ \biggl( {\frac {919773569303}{279417600}}+{\frac {
5171}{36}}\,{\pi }^{2}+{\frac {2106081}{448}}\,\ln  \left( 3 \right) -
{\frac {12295049}{1260}}\,\ln  \left( 2 \right) 
\\ \nonumber && \hspace{1.0cm}
-{\frac {553297}{1260}}\,\gamma
+{\frac {208571}{1792}}\,{q}^{2}-{\frac {42271}{96}}\,\pi \,Y
q+{\frac {471711}{1792}}\,{Y}^{2}{q}^{2} \biggr) {e}^{4}
\\ \nonumber && \hspace{0.5cm}
+ \biggl( {
\frac {308822406727}{186278400}}-{\frac {864819261}{35840}}\,\ln 
 \left( 3 \right) -{\frac {187357}{1260}}\,\gamma+{\frac {1751}{36}}\,
{\pi }^{2}
\\ \nonumber && \hspace{1.0cm}
-{\frac {5224609375}{193536}}\,\ln  \left( 5 \right) 
+{\frac {24908851}{252}}\,\ln  \left( 2 \right) +{\frac {3253}{10752}}\,
{q}^{2}-{\frac {4867907}{27648}}\,\pi \,Yq
\\ \nonumber && \hspace{1.0cm}
+{\frac {289063}{1536}}\,{Y}
^{2}{q}^{2} \biggr) {e}^{6}
\\ && \hspace{0.5cm}
- \left( {\frac {1712}{105}}+{\frac {14552}{63}}\,{e}^{2}+{\frac {553297}{1260}}\,{e}^{4}+{\frac {187357}{1260}}\,{e}^{6} \right) \ln v \biggr\} {v}^{6}
\biggr], \label{eq:dotE8}\\
\left\langle \frac{dL}{dt} \right\rangle_t^\infty &=&
\left( \frac{dL}{dt} \right)_{\rm N}
\biggl[ \left\{1+{\frac {7}{8}}\,{e}^{2}\right\}\,Y
+ \left\{ 
-{\frac {1247}{336}}-{\frac {425}{336}}\,{e}^{2}+{\frac {10751}{2688}}\,{e}^{4}
\right\} \,Y\,{v}^{2}
\nonumber \\ \nonumber && 
+ \biggl\{ 
{\frac {61}{24}}\,q-{\frac {61}{8}}\,{Y}^{2}q+4\,\pi \,Y 
+ \left( {\frac {63}{8}}\,q+{\frac {97}{8}}\,\pi \,Y-{\frac {91}{4}}
\,{Y}^{2}q \right) {e}^{2}
\\ \nonumber && \hspace{0.5cm}
+\left( {\frac {95}{64}}\,q+{\frac {49}{32}}\,\pi \,Y-{\frac {461}{64}}\,{Y}^{2}q \right) {e}^{4}
-{\frac {49}{4608}}\,\pi \,Y{e}^{6}
\biggr\} {v}^{3}
\\ \nonumber && 
+ \biggl\{ -{\frac {44711}{9072}}-{\frac {57}{16}}{q}^{2}+{\frac {45}{8}}\,{Y}^{2}{q}^{2}+ \left( -{\frac {302893}{6048}}-{\frac {201}{16}}{q}^{2}+{\frac {37}{2}}\,{Y}^{2}{q}^{2} \right) {e}^{2}
\\ \nonumber && \hspace{0.5cm}
+ \left( -{\frac {701675}{24192}}-{
\frac {351}{128}}{q}^{2}+{\frac {331}{64}}\,{Y}^{2}{q}^{2} \right) {e}^{4}
+{\frac {162661}{16128}}{e}^{6} \biggr\}\,Y\,{v}^{4}
\\ \nonumber && 
+ \biggl\{ 
{\frac {4301}{224}}\,{Y}^{2}q-{\frac {8191}{672}}\,\pi \,Y-{\frac {2633}{224}}\,q 
\\ \nonumber && \hspace{0.5cm}
+ \left( -{\frac {66139}{1344}}\,q-{\frac {48361}{1344}}\,\pi \,Y+{\frac {18419}{448}}\,{Y}^{2}q \right) {e}^{2}
\\ \nonumber && \hspace{0.5cm}
+ \left( {\frac {3959}{1792}}\,q+{\frac {1657493}{43008}}\,\pi \,Y-{\frac {257605}{5376}}\,{Y}^{2}q \right) {e}^{4}
\\ \nonumber && \hspace{0.5cm}
+ \left( {\frac {19161}{3584}}\,q+{\frac {5458969}{774144}}\,\pi \,Y-{
\frac {52099}{1536}}\,{Y}^{2}q \right) {e}^{6}\biggr\} {v}^{5}
\\ \nonumber && 
+ \biggl\{ {\frac {145}{12}}\,\pi \,q+{\frac {
6643739519}{69854400}}\,Y+\frac{16}{3}\,{\pi }^{2}Y-{\frac {1712}{105}}\,
\gamma\,Y-{\frac {3424}{105}}\,\ln  \left( 2 \right) Y
\\ \nonumber && \hspace{0.5cm}
-{\frac {171}{112}}\,Y{q}^{2}
-{\frac {145}{4}}\,\pi \,{Y}^{2}q+{\frac {1769}{112}}\,
{Y}^{3}{q}^{2}
\\ \nonumber && \hspace{0.5cm}
+ \biggl( {\frac {995}{12}}\,\pi \,q+{\frac {229}{6}}\,{
\pi }^{2}Y+{\frac {6769212511}{8731800}}\,Y+{\frac {1391}{30}}\,\ln 
 \left( 2 \right) Y-{\frac {24503}{210}}\,\gamma\,Y
\\ \nonumber && \hspace{1.0cm}
-{\frac {78003}{280
}}\,\ln  \left( 3 \right) Y-{\frac {46867}{1344}}\,Y{q}^{2}-{\frac {
877}{4}}\,\pi \,{Y}^{2}q+{\frac {27997}{192}}\,{Y}^{3}{q}^{2} \biggr) 
{e}^{2}
\\ \nonumber && \hspace{0.5cm}
+ \biggl( {\frac {21947}{384}}\,\pi \,q+{\frac {4795392143}{
7761600}}\,Y+{\frac {3042117}{1120}}\,\ln  \left( 3 \right) Y-{\frac {
418049}{84}}\,\ln  \left( 2 \right) Y
\\ \nonumber && \hspace{1.0cm}
-{\frac {11663}{140}}\,\gamma\,Y
+{\frac {109}{4}}\,{\pi }^{2}Y-{\frac {1481}{16}}\,Y{q}^{2}-{\frac {
22403}{128}}\,\pi \,{Y}^{2}q+{\frac {267563}{1344}}\,{Y}^{3}{q}^{2}
 \biggr) {e}^{4}
\\ \nonumber && \hspace{0.5cm}
+ \biggl( {\frac {38747}{13824}}\,\pi \,q+{\frac {
31707715321}{186278400}}\,Y+{\frac {23}{16}}\,{\pi }^{2}Y+{\frac {
94138279}{2160}}\,\ln  \left( 2 \right) Y
\\ \nonumber && \hspace{1.0cm}
-{\frac {1044921875}{96768}}
\,\ln  \left( 5 \right) Y-{\frac {42667641}{3584}}\,\ln  \left( 3
 \right) Y-{\frac {2461}{560}}\,\gamma\,Y-{\frac {68333}{3584}}\,Y{q}^{2}
\\ \nonumber && \hspace{1.0cm}
-{\frac {59507}{4608}}\,\pi \,{Y}^{2}q+{\frac {183909}{3584}}\,{Y}^
{3}{q}^{2} \biggr) {e}^{6}
\\ && \hspace{0.5cm}
- \left( {\frac {1712}{105}}+{\frac {24503}{210}}{e}^{2}+{\frac {11663}{140}}{e}^{4}+{\frac {2461}{560}}{e}^{6} \right) \,Y \ln v \biggr\} {v}^{6}
\biggr],\label{eq:dotL8}\\
\left\langle \frac{dC}{dt} \right\rangle_t^\infty &=&
\left( \frac{dC}{dt} \right)_{\rm N}
\biggl[
1+{\frac {7}{8}}\,{e}^{2}+ \left( -{\frac {743}{336}}+{\frac {23}{42}}
\,{e}^{2}+{\frac {11927}{2688}}\,{e}^{4} \right) {v}^{2}
\nonumber \\ \nonumber && 
+ \biggl\{ 4\,\pi-{\frac {85}{8}}\,Yq
 + \left( {\frac {97}{8}}\,\pi -{\frac {211}{
8}}\,Yq \right) {e}^{2}
\\ \nonumber && \hspace{0.5cm}
+ \left( {\frac {49}{32}}\,\pi -{\frac {517}{64
}}\,Yq \right) {e}^{4}-{\frac {49}{4608}}\,\pi \,{e}^{6} \biggr\} {v}^{3}
\\ \nonumber &&
+ \biggl\{ -{\frac {129193}{18144}}-{\frac {329}{96}}\,{q}^{2}+{\frac 
{53}{8}}\,{Y}^{2}{q}^{2}+ \left( -{\frac {84035}{1728}}-{\frac {929}{
96}}\,{q}^{2}+{\frac {163}{8}}\,{Y}^{2}{q}^{2} \right) {e}^{2}
\\ \nonumber && \hspace{0.5cm}
+ \left( -{\frac {1030273}{48384}}-{\frac {1051}{768}}\,{q}^{2}+{\frac 
{387}{64}}\,{Y}^{2}{q}^{2} \right) {e}^{4}+{\frac {100103}{8064}}\,{e}
^{6} \biggr\} {v}^{4}
\\ \nonumber && 
+ \biggl\{ -{\frac {4159}{672}}\,\pi +{\frac {2553}{224}}\,Yq
+ \left( -{\frac {21229}{1344}}\,\pi -{\frac {553}{192}}\,Yq
 \right) {e}^{2}\\ \nonumber && \hspace{0.5cm}
+ \left( {\frac {2017013}{43008}}\,\pi -{\frac {475541
}{5376}}\,Yq \right) {e}^{4}+ \left( {\frac {6039325}{774144}}\,\pi -{
\frac {153511}{3584}}\,Yq \right) {e}^{6} \biggr\} {v}^{5}
\\ \nonumber && 
+ \biggl\{ 
{\frac {11683501663}{139708800}}+\frac{16}{3}\,{\pi }^{2}
-{\frac {1712}{105}}\,\gamma-{\frac {3424}{105}}\,\ln  \left( 2 \right) 
+{\frac {1277}{192}}\,{q}^{2}-{\frac {193}{4}}\,\pi \,Yq
\\ \nonumber && \hspace{0.5cm}
+{\frac {2515}{48}}\,{Y}^{2}{q}^{2}
+ \biggl( {\frac {16319179321}{23284800}}+{\frac {229}{6}}\,{\pi }^{2}
-{\frac {24503}{210}}\,\gamma+{\frac {1391}{30}}\,\ln  \left( 2
 \right) 
\\ \nonumber && \hspace{1.0cm}
-{\frac {78003}{280}}\,\ln  \left( 3 \right) 
+{\frac {16979}{1344}}\,{q}^{2}-{\frac {2077}{8}}\,\pi \,Yq+{\frac {118341}{448}}\,{Y}^{2}{q}^{2} \biggr) {e}^{2}
\\ \nonumber && \hspace{0.5cm}
+ \biggl( {\frac {211889615389}{372556800}}+
{\frac {109}{4}}\,{\pi }^{2}+{\frac {3042117}{1120}}\,\ln  \left( 3
 \right) -{\frac {11663}{140}}\,\gamma-{\frac {418049}{84}}\,\ln 
 \left( 2 \right) 
\\ \nonumber && \hspace{1.0cm}
-{\frac {132193}{3584}}\,{q}^{2}-{\frac {24543}{128}
}\,\pi \,Yq+{\frac {91747}{336}}\,{Y}^{2}{q}^{2} \biggr) {e}^{4}
\\ \nonumber && \hspace{0.5cm}
+ \biggl( {\frac {33928992071}{186278400}}-{\frac {1044921875}{96768}}\,
\ln  \left( 5 \right) +{\frac {23}{16}}\,{\pi }^{2}-{\frac {42667641}{
3584}}\,\ln  \left( 3 \right) 
\\ \nonumber && \hspace{1.0cm}
+{\frac {94138279}{2160}}\,\ln  \left( 2 \right) 
-{\frac {2461}{560}}\,\gamma-{\frac {24505}{5376}}\,{q}^{2}-{
\frac {4151}{288}}\,\pi \,Yq+{\frac {718799}{10752}}\,{Y}^{2}{q}^{2}
 \biggr) {e}^{6}
\\ && \hspace{0.5cm}
- \left( {\frac {1712}{105}}+{\frac {24503}{210}}\,{e}^{2}+{\frac {11663}{140}}\,{e}^{4}+{\frac {2461}{560}}\,{e}^{6}
 \right) \ln v \biggr\} {v}^{6}
\biggl],\label{eq:dotC8}
\end{eqnarray}
where the leading contributions are given by
\begin{eqnarray}
\left( \frac{dE}{dt} \right)_{\rm N} &=&
- \frac{32}{5} \left( \frac{\mu}{M} \right)^2
v^{10} (1-e^2)^{3/2},
\nonumber \\
\left( \frac{dL}{dt} \right)_{\rm N} &=&
- \frac{32}{5} \left( \frac{\mu^2}{M} \right)
v^{7} (1-e^2)^{3/2},
\nonumber \\
\left( \frac{dC}{dt} \right)_{\rm N} &=&
- \frac{64}{5} \mu^3 v^{6}
 (1-e^2)^{3/2} (1-Y^2).
\label{eq:dotFN}
\end{eqnarray}

The horizon parts of Eqs.(\ref{eq:Edot})-(\ref{eq:Cdot})
are given by 
\begin{eqnarray}
\left\langle \frac{dE}{dt} \right\rangle_t^{\rm H} &=&
\left( \frac{dE}{dt} \right)_{\rm N}
\biggl[
-{\frac {1}{512}}\, \left( 16+120\,{e}^{2}+90\,{e}^{4}+5\,{e}^{6} 
 \right)  \left( 8+9\,{q}^{2}+15\,{Y}^{2}{q}^{2} \right) 
q Y v^5
\nonumber \\ &&
-\biggl\{ 
1+{\frac {81}{32}}\,{q}^{2}-{\frac {15}{32}}\,{Y}^{2}{q}^{2}+ \left( {
\frac {57}{4}}+{\frac {1143}{32}}\,{q}^{2}-{\frac {195}{32}}\,{Y}^{2}{
q}^{2} \right) {e}^{2}
\nonumber \\  && \hspace{0.5cm}
+ \left( {\frac {465}{16}}+{\frac {4455}{64}}\,{
q}^{2}-{\frac {225}{32}}\,{Y}^{2}{q}^{2} \right) {e}^{4}
\nonumber \\  && \hspace{0.5cm}
+ \left( {
\frac {355}{32}}+{\frac {6345}{256}}\,{q}^{2}+{\frac {75}{256}}\,{Y}^{
2}{q}^{2} \right) {e}^{6}
\biggr\} q Y v^7
\biggr], \label{eq:dotEH} \\
\left\langle \frac{dL}{dt} \right\rangle_t^{\rm H} &=&
\left( \frac{dL}{dt} \right)_{\rm N}
\biggl[
-{\frac {\left( 8+24\,{e}^{2}+3\,{e}^{4} \right)}{1024}}   \left( 
16+33\,{q}^{2}+16\,{Y}^{2}+18\,{Y}^{2}{q}^{2}+45\,{Y}^{4}{q}^{2}
 \right) q v^5
\nonumber \\  && 
-\biggl\{
\frac{5}{4}+{\frac {375}{128}}\,{q}^{2}-\frac{1}{4}\,{Y}^{2}
-{\frac {63}{64}}\,{Y}^{2}{q}^{2}+{\frac {15}{128}}\,{Y}^{4}{q}^{2}
\nonumber \\  && \hspace{0.5cm}
+ \left( 10+{\frac {5955}{256}}\,{q}^{2}-\frac{5}{4}\,{Y}^{2}-{\frac {855}{128}}\,{Y}^{2}{q}^{2}+{\frac {675}{256}}\,{Y}^{4}{q}^{2} \right) {e}^{2}
\nonumber \\   && \hspace{0.5cm}
+ \left( {\frac {255}{32}}+{\frac {18855}{1024}}\,{q}^{2}-{\frac {15}{32}}\,{Y}^{2}-{\frac {2295}{512}}\,{Y}^{2}{q}^{2}+{\frac {3375}{1024}}\,{Y}^{4}{q}^{2} \right) {e}^{4}
\nonumber \\  && \hspace{0.5cm}
+ \left( {\frac {15}{32}}+{\frac {2205}{2048}}\,{q}^{2}-{\frac {225}{1024}}\,{Y}^{2}{q}^{2}+{\frac {525}{2048}}\,{Y}^{4}{q}^{2}
 \right) {e}^{6}\biggr\}
q v^7
\biggl], \label{eq:dotLH} \\
\left\langle \frac{dC}{dt} \right\rangle_t^{\rm H} &=&
\left( \frac{dC}{dt} \right)_{\rm N}
\biggl[
-{\frac {1}{1024}} \left( 8+24\,{e}^{2}+3\,{e}^{4} \right) 
 \left( 16+3\,{q}^{2}+45\,{Y}^{2}{q}^{2} \right) q Y v^5 
\nonumber \\ && 
+ 
\biggl\{
{\frac{1}{16}+\frac {93}{256}}\,{q}^{2}-{\frac {165}{256}}\,{Y}^{2}{q}^{2}+
 \left(\frac{5}{8}+{\frac {705}{256}}\,{q}^{2}-{\frac {1125}{256}}\,{Y}^{2}{q
}^{2} \right) {e}^{2}
\nonumber \\ && \hspace{0.5cm}
+ \left( {\frac {27}{128}}+{\frac {4131}{2048}}\,
{q}^{2}-{\frac {8235}{2048}}\,{Y}^{2}{q}^{2} \right) {e}^{4}
\nonumber \\ && \hspace{0.5cm}
+ \left( -
{\frac {3}{128}}+{\frac {27}{256}}\,{q}^{2}-{\frac {165}{512}}\,{Y}^{2
}{q}^{2} \right) {e}^{6}
\biggr\} q Y v^7
\biggr]. \label{eq:dotCH}
\end{eqnarray}
$\langle dE/dt \rangle^{\rm H}_t$, $\langle dL/dt \rangle^{\rm H}_t$
and $\langle dC/dt \rangle^{\rm H}_t$ 
in Eqs.~(\ref{eq:dotEH})-(\ref{eq:dotCH}) are new PN formulae 
derived in this paper. $\langle dE/dt \rangle^\infty_t$,
$\langle dL/dt \rangle^\infty_t$ and $\langle dC/dt \rangle^\infty_t$
in Eqs.~(\ref{eq:dotE8})-(\ref{eq:dotC8}) are consistent with those
in Ref.~\cite{Ganz:2007rf} up to 2.5PN and $O(e^2)$.

From the leading order expressions in Eq.~(\ref{eq:dotFN}), 
one will find that the Carter parameter, $C$,
does not change due to the radiation of the gravitational waves 
when $Y=1$ (equatorial orbits) because $(dC/dt)_{\rm N}=0$.

In the Schwarzschild case, the Carter parameter corresponds to the
square of the equatorial angular momentum (the normal component to the
rotational axis of the central black hole). Then there is expected
to exist the duality between $L^2$ and $C$ due to the spherical symmetry.
In fact, from Eqs.~(\ref{eq:dotL8}) and (\ref{eq:dotC8}),
(and also from (\ref{eq:dotLH}) and (\ref{eq:dotCH})),
one can find that $\langle dL^2/dt \rangle_t$ vanishes in $Y=0$
(polar orbits) while $\langle dC/dt \rangle_t$ for $Y=0$ coincides with
$\langle dL^2/dt \rangle_t$ for $Y=1$.
This can be also realized by seeing that the secular change of the total
angular momentum, $\langle d(L^2+C)/dt \rangle_t$, is independent of $Y$.
Then, it might be possible to understand that 
$\langle dL/dt \rangle_t^\infty$ becomes 1.5PN from the leading order 
when $q\neq 0$ and $Y=0$ (polar orbits) due to the spin-orbit coupling.

From the expressions of the horizon parts shown in
Eqs.~(\ref{eq:dotEH})-(\ref{eq:dotCH}),
we find that the absorption of the gravitational waves to the central
black hole contributes at $O(v^5)$ from the leading order
in Eq.~(\ref{eq:dotFN}) for $q\ne0$ and at $O(v^8)$ for $q=0$.
The $O(v^5)$ and $O(v^7)$ corrections in
$\langle dE/dt \rangle_t^{\rm H}$
can be positive for $q>0$, which means that the particle can gain
the energy through a superradiance phenomenon.
These observations are consistent with the results
for circular, equatorial orbits
shown in Refs.~\cite{Poisson:1994yf,Tagoshi:1997jy}.

We also find that the superradiance terms in Eq.~(\ref{eq:dotEH})
vanish for $Y=0$, and that
$\langle dE/dt \rangle_t^{\rm H}$ has only the 4PN and higher order
corrections. The superradiance terms may come from the coupling
between the black hole spin and the orbital angular momentum, like
$\propto \textbf{L}\cdot\textbf{S} \propto q\cos\iota$. Hence, when
the orbital inclination increases ($Y$ gets small correspondingly),
the superradiance is suppressed \cite{Hughes:2001jr}.

\subsection{Comparison to numerical results} \label{sec:compare}

To investigate the accuracy of the 4PN $O(e^6)$ formulae derived
in this work, we compare them to the
corresponding numerical results given by the method established in 
Ref.~\cite{Fujita:2004rb, Fujita:2009uz, Fujita:2009us}, 
which enables one to compute the modal fluxes with the relative error of 
$\sim 10^{-14}$ in double precision computations.
In the practical computations, as well as in deriving the analytic
expressions, we need to truncate the summation to finite ranges of
$\tilde\Lambda=\{\ell,m,n_r,n_\theta\}$
in Eqs.~(\ref{eq:Edot})-(\ref{eq:Cdot}).
In order to save the computation time in the numerical calculation, 
we sum $\ell$ up to $7$. 
We can check that the error due to neglecting terms for $\ell\ge 8$ 
is smaller than the relative error in the 4PN $O(e^6)$ formulae 
from the corresponding numerical results up to $\ell=7$. 
We also truncate $n_r$ and $n_\theta$ to achieve the relative error 
of $\sim 10^{-7}$ in numerical results up to $\ell=7$. 
For the parameters investigated in the comparison, 
the relative error of $\sim 10^{-7}$ achieved 
by truncating $n_r$ and $n_\theta$ is again smaller than the relative error 
in the 4PN $O(e^6)$ formulae from the numerical results up to $\ell=7$. 
Thus, we can regard numerical results as benchmarks to 
investigate the accuracy in our analytic formulae. 

Here we define the relative error in the analytic formula of
$\langle dE/dt \rangle_t$ by
\begin{equation}
\Delta_{E} \equiv
\left|
1 - \left\langle\frac{dE}{dt}\right\rangle_t^{\rm Ana}
\bigg/ \left\langle\frac{dE}{dt}\right\rangle_t^{\rm Num}
\right|, \label{eq:relative_error}
\end{equation}
where $\left\langle{dE}/{dt}\right\rangle_t^{\rm Ana}$ denotes
the analytic formula in order to distinguish it from the corresponding
numerical result, $\left\langle{dE}/{dt}\right\rangle_t^{\rm Num}$.
We also define the relative errors in the analytic formulae of
$\langle dL/dt \rangle_t$ and $\langle dC/dt \rangle_t$ in a similar
manner and denote them as
$\Delta_L$ and $\Delta_C$ respectively.

\begin{figure}
\includegraphics[width=51mm]{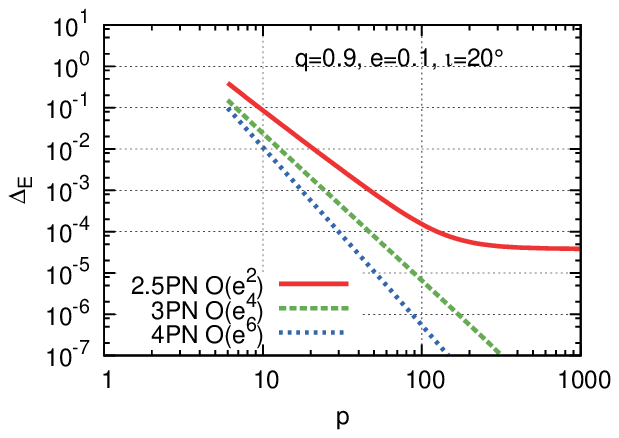}%
\includegraphics[width=51mm]{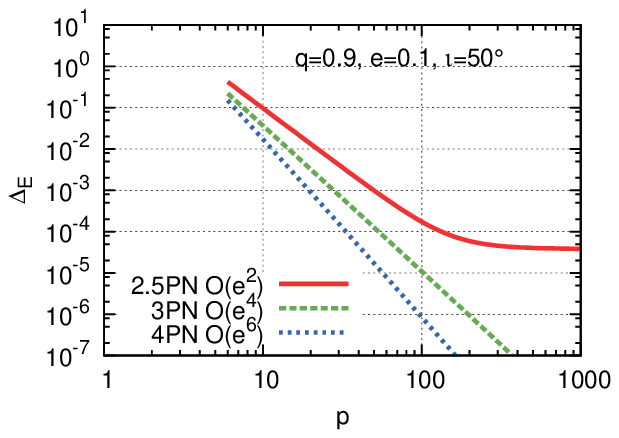}%
\includegraphics[width=51mm]{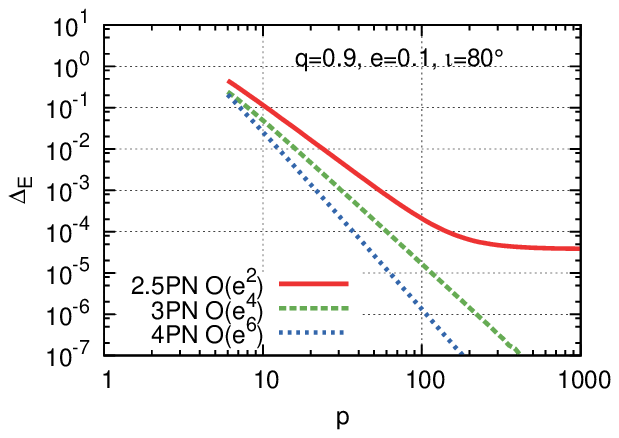}\\
\includegraphics[width=51mm]{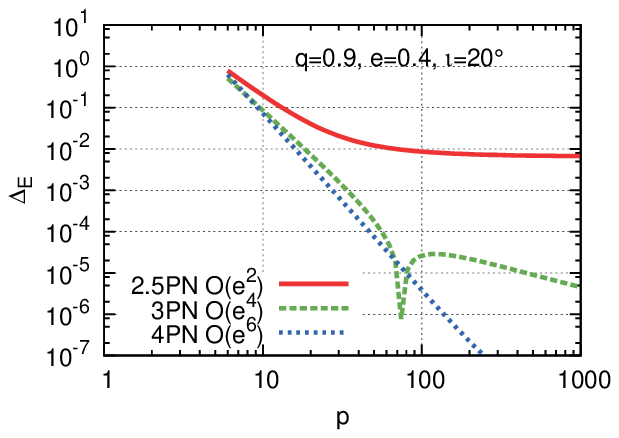}%
\includegraphics[width=51mm]{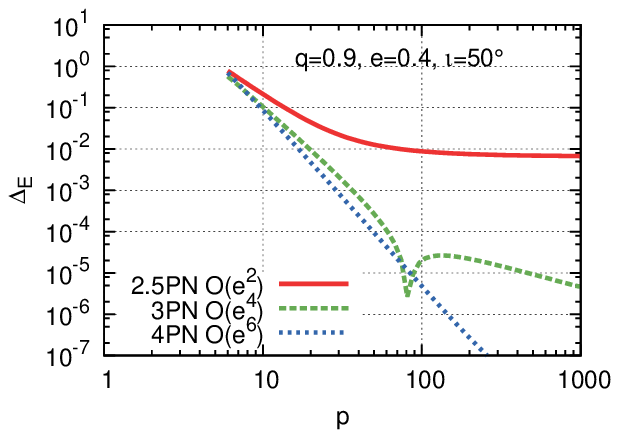}%
\includegraphics[width=51mm]{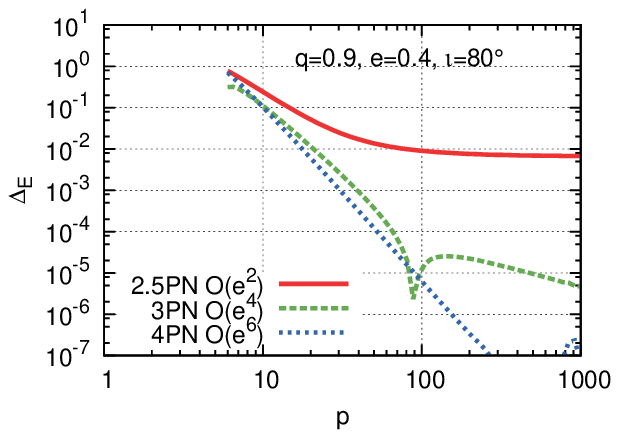}\\
\includegraphics[width=51mm]{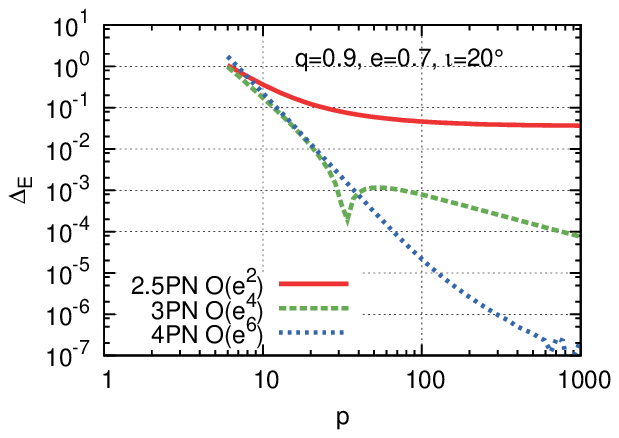}%
\includegraphics[width=51mm]{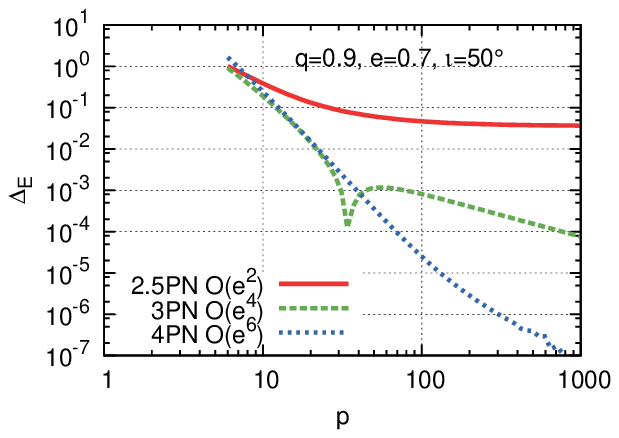}%
\includegraphics[width=51mm]{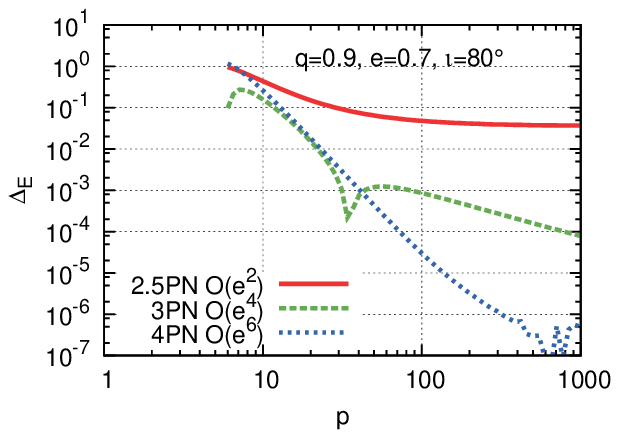}
\caption{The relative errors in the analytic PN formulae for
the secular change of the particle's energy as a function of 
the semi-latus rectum $p$ for $q=0.9$, $e=0.1, 0.4$ and $0.7$ 
(from top to bottom) and $\iota=20^{\circ}, 50^{\circ}$ and $80^{\circ}$ 
(from left to right).
In addition to the error in the 4PN $O(e^6)$ formula, those in the 2.5PN
$O(e^2)$ and the 3PN $O(e^4)$ formulae are shown in each plot for reference.
We truncated the plots at $p=6$ because the relative errors get too large
(nearly or more than unity) in $p<6$ to be meaningful.
One finds that the relative error becomes smaller with increasing orders of 
the PN approximation and the expansion with respect to the eccentricity. 
The relative error in our 4PN $O(e^6)$ formula falls off faster than $p^{-4}$ 
when the eccentricity is small, {\it e.g.} $e\lesssim 0.4$. 
Since $v=\sqrt{1/p}$, this would imply that our 4PN formula 
is correctly representing the secular change up to the 4PN order. 
Note, however, that the relative error in the 4PN $O(e^6)$ formula 
for $e=0.7$ falls off slower than $p^{-4}$ when the semi-latus rectum
becomes larger,{\it e.g.} $p>100$.
This might be because of the higher order corrections of $e$ than $O(e^6)$,
which will contain the lower PN terms than the 4PN order.
We also note that changing the inclination angle, $\iota$, 
does not change the dependence on $p$ of the relative error
for fixed $q$ and $e$ so much.
This might be checked more easily in contour plots in
Fig.~\ref{fig:dotE8H_q0.9_m0.9_inc20_80}, which show
the relative error as a function of $p$ and $e$ for fixed $q$ and $\iota$. 
} \label{fig:dotE8H_q0.9_e0.1_0.7_inc20_80}
\end{figure}

Fig.~\ref{fig:dotE8H_q0.9_e0.1_0.7_inc20_80} shows several plots
of $\Delta_E$ for the 4PN $O(e^6)$ formula as a function of $p$ for
several sets of $(e,\iota)$ with $q=0.9$. In the plots,
we also show the relative
errors in the 2.5PN $O(e^2)$ and 3PN $O(e^4)$ formulae for reference.
From the plots for $e=0.1$ (three on the top), one can find that
$\Delta_E$ falls off faster than $p^{-4}$ for
$p \gtrsim 10$ (Similarly, the relative errors in the 2.5PN $O(e^2)$
and 3PN $O(e^4)$ formulae fall off faster than $p^{-5/2}$ and
$p^{-3}$).
Noting $v=\sqrt{1/p}$, this would be a good indication that 
our PN formula has been derived correctly up to required order.

$\Delta_E$ is expected to contain not only higher order corrections
than the 4PN order, but also the higher order corrections of eccentricity
than $O(e^6)$ in the lower PN terms, which will become dominant when
$p$ and $e$ get larger.
In fact, seeing the plots for $e=0.7$ in
Fig.~\ref{fig:dotE8H_q0.9_e0.1_0.7_inc20_80},
one can find that the relative error strays out of the expected
power law line for large $p$.
This behavior is clearer in the plots of the relative error in the
2.5PN $O(e^2)$ formula. From Eqs.~(\ref{eq:dotE8}) and (\ref{eq:dotEH}),
we know that the relative error in the 2.5PN $O(e^2)$ formula contains
the $O(e^4)$ correction in the $O(v^0)$ term. The effect of this
correction appears as large-$p$ plateaus in the plots
(also see Fig.~\ref{fig:dotE_en_q0.9_Y1_0}).
This may motivate us to perform the higher order expansion
with respect to the orbital eccentricity in the PN formulae 
or to derive the PN formulae without performing 
the expansion with respect to the orbital eccentricity
\cite{Galtsov:1980ef,Shibata:1994xk,Arun:2007sg,Arun:2009mc}.
In addition, it might be noted that the behavior of the relative error
does not strongly depend on the inclination angle $\iota$ for fixed $q$
and $e$.

\begin{figure}
\includegraphics[width=51mm]{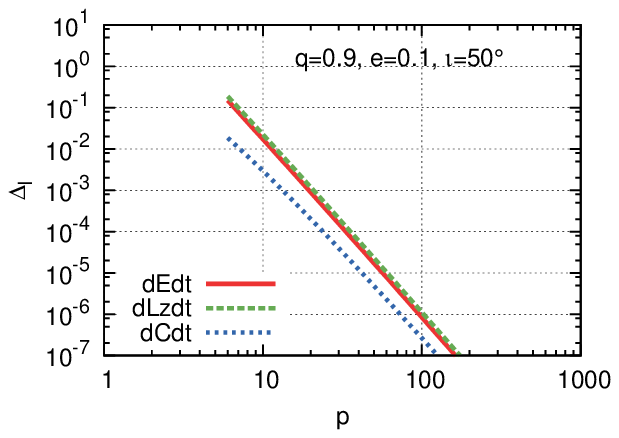}%
\includegraphics[width=51mm]{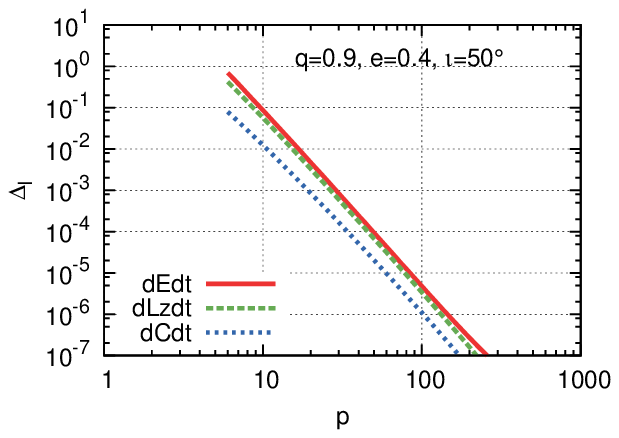}%
\includegraphics[width=51mm]{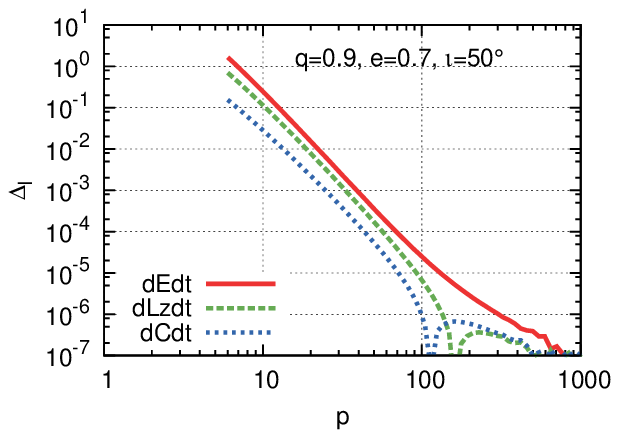}\\
\includegraphics[width=51mm]{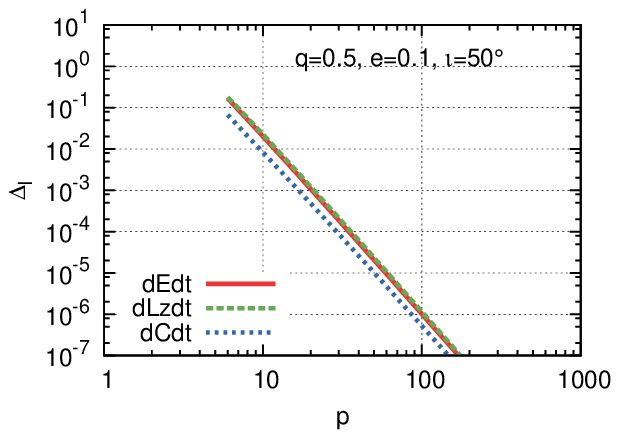}%
\includegraphics[width=51mm]{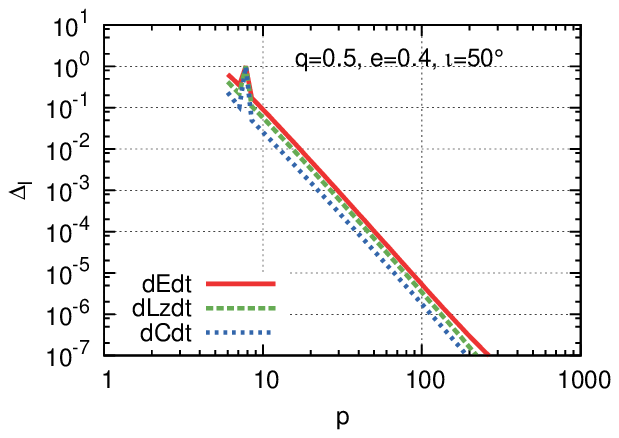}%
\includegraphics[width=51mm]{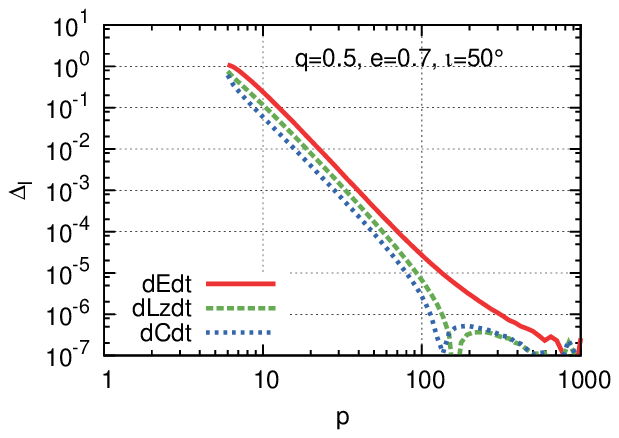}\\
\includegraphics[width=51mm]{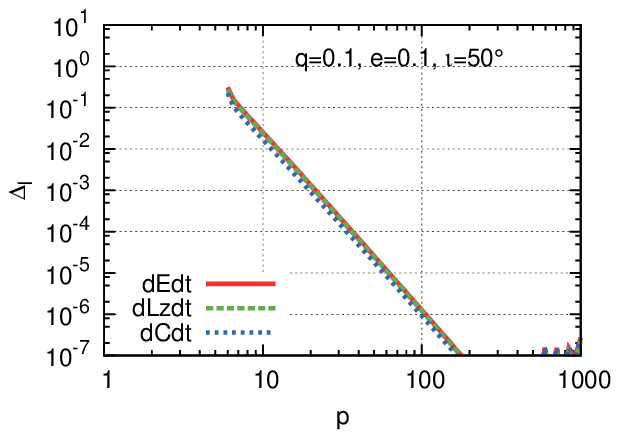}%
\includegraphics[width=51mm]{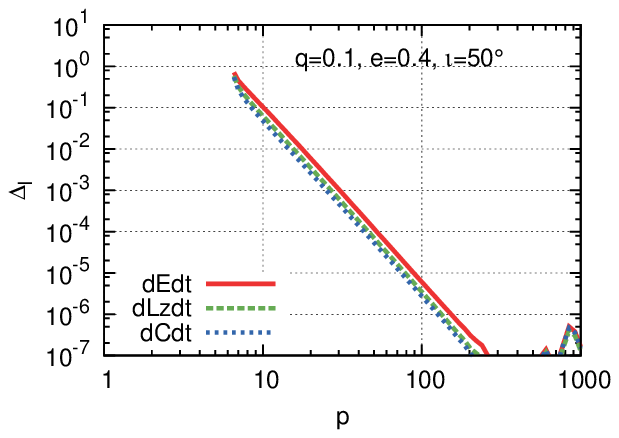}%
\includegraphics[width=51mm]{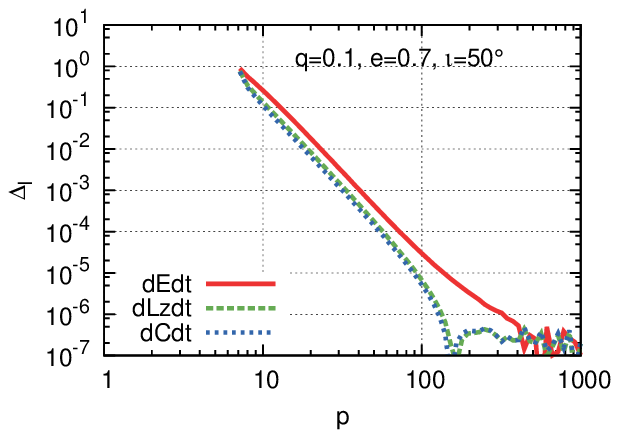}\\
\includegraphics[width=51mm]{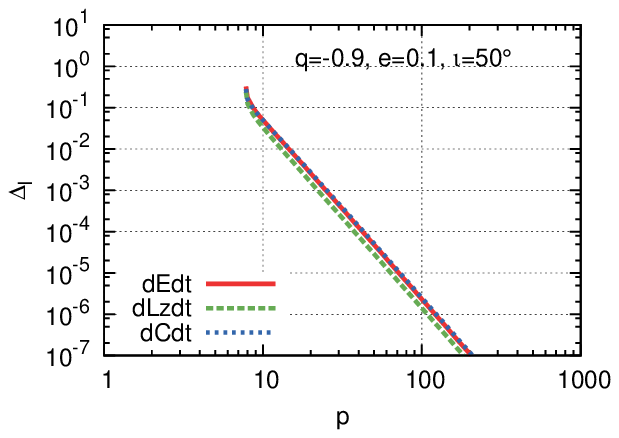}%
\includegraphics[width=51mm]{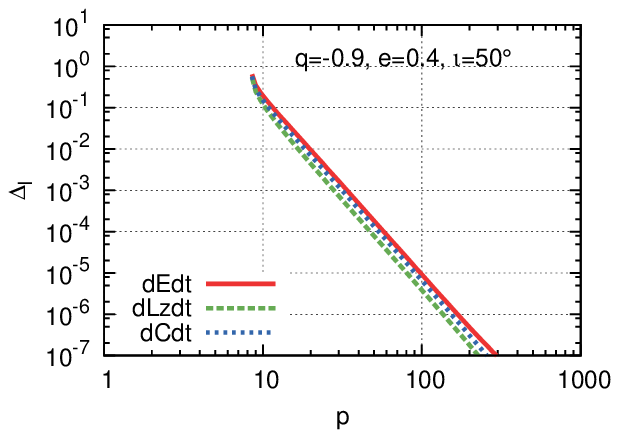}%
\includegraphics[width=51mm]{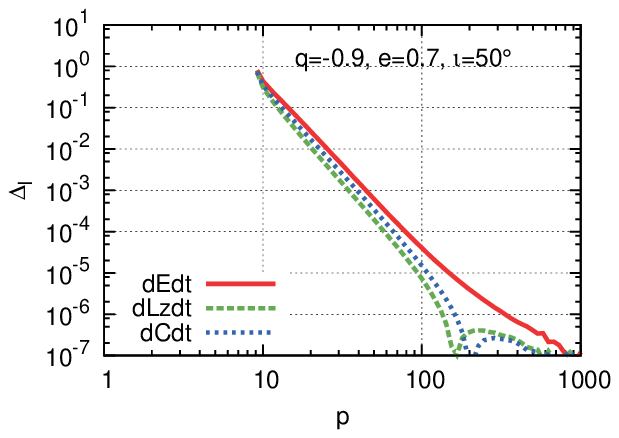}
\caption{The relative errors in the analytic PN formulae for the secular
changes of the three orbital parameters, $\{E,L,C\}$, as functions of 
the semi-latus rectum $p$ for $\iota=50^{\circ}$, $q=0.9, 0.5, 0.1$ and $-0.9$
(from top to bottom) and $e=0.1, 0.4$ and $0.7$ (from left to right). 
We truncated the plots at $p=\max\{6, p_s(e,\iota)\}$, where $p_s(e,\iota)$
is the value of $p$ at the ``separatrix'' (the boundary between stable and
unstable orbits), because the relative errors get too large in
$p<6$ to be meaningful and the orbit is not stable for $p<p_s(e,\iota)$.
As pointed out in Fig.~\ref{fig:dotE8H_q0.9_e0.1_0.7_inc20_80}, 
the relative errors in our 4PN $O(e^6)$ formulae fall off faster than
$p^{-4}$ when the eccentricity is small, {\it e.g.} $e\lesssim 0.4$,
while the fall-off gets slower when $p$ is larger for $e=0.7$.
There are not large differences in the behaviors of $\Delta_E$, $\Delta_L$
and $\Delta_C$. This suggests that
it might be enough to focus only on $\langle dE/dt\rangle_t$ to
investigate the accuracy and convergence of our 4PN formulae.
} \label{fig:dotF8H_q0.9_m0.9_e0.1_0.4_0.7_inc50}
\end{figure}

In Fig.~\ref{fig:dotF8H_q0.9_m0.9_e0.1_0.4_0.7_inc50}, 
we show the relative errors in the 4PN $O(e^6)$ formulae for the secular
changes of the three orbital parameters, $\{E,L,C\}$,
for several sets of $(q,e)$ and $\iota=50^{\circ}$.
As in the case of $\Delta_E$
shown in Fig.~\ref{fig:dotE8H_q0.9_e0.1_0.7_inc20_80}, 
the relative errors, $\Delta_L$ and $\Delta_C$, fall off faster
than $p^{-4}$ when $p\gtrsim 10$, except for the large $p$ region
($p \gtrsim 100$) in the case of $e=0.7$.
Thus, the 4PN $O(e^6)$ formulae for the secular changes of the orbital
parameters are expected to be valid up to $O(v^8)$.
From Fig.~\ref{fig:dotF8H_q0.9_m0.9_e0.1_0.4_0.7_inc50}, 
one might think that it is enough to investigate only $\Delta_{E}$ 
to discuss the accuracy of our formulae
since there are not large differences in the relative errors,
$\Delta_E$, $\Delta_L$ and $\Delta_C$.

\begin{figure}
\includegraphics[width=51mm]{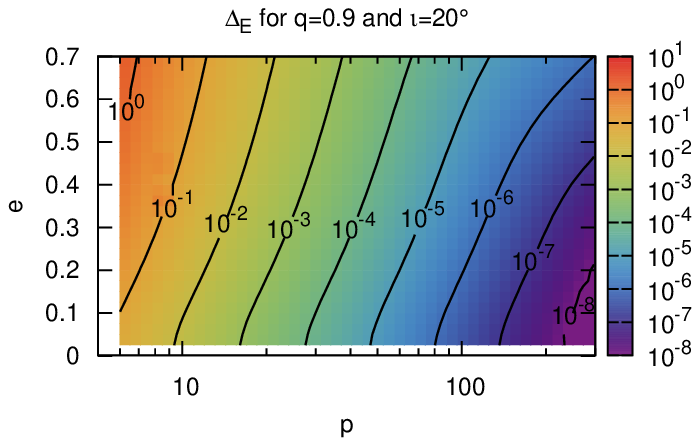}%
\includegraphics[width=51mm]{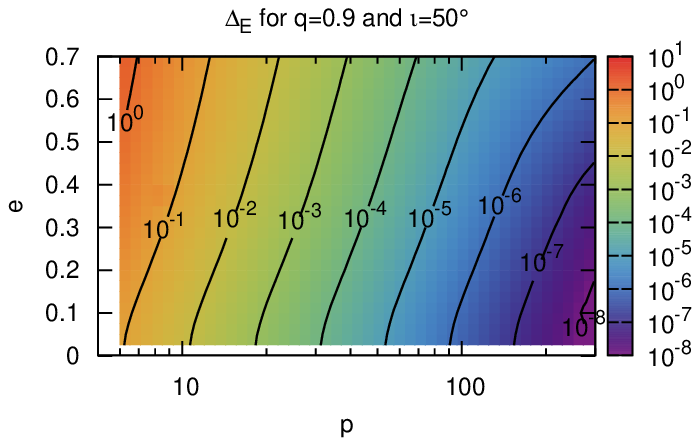}%
\includegraphics[width=51mm]{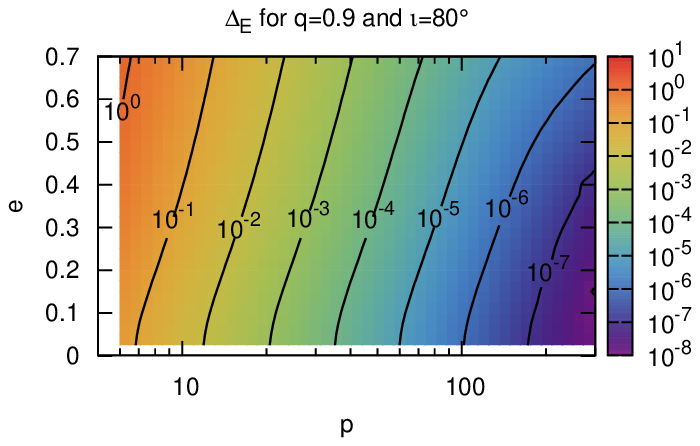}\\
\includegraphics[width=51mm]{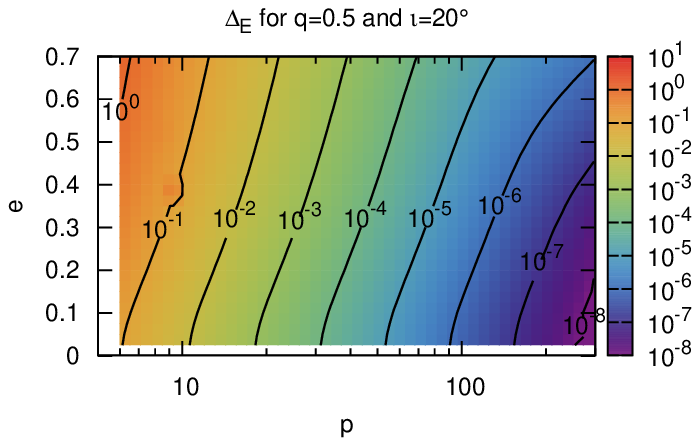}%
\includegraphics[width=51mm]{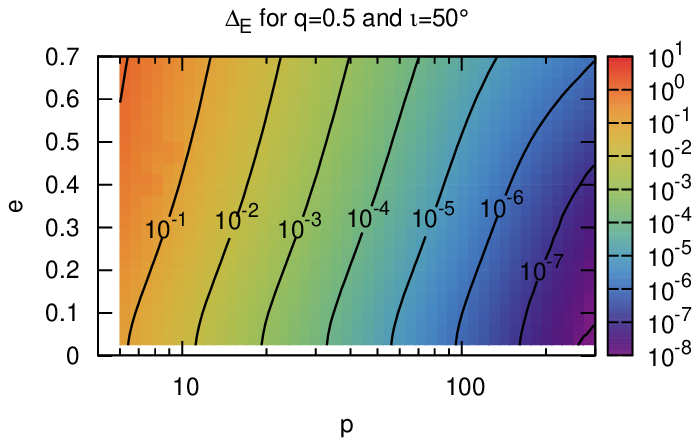}%
\includegraphics[width=51mm]{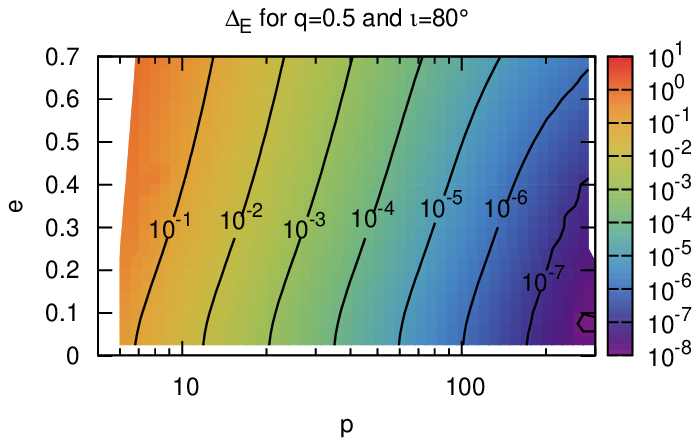}\\
\includegraphics[width=51mm]{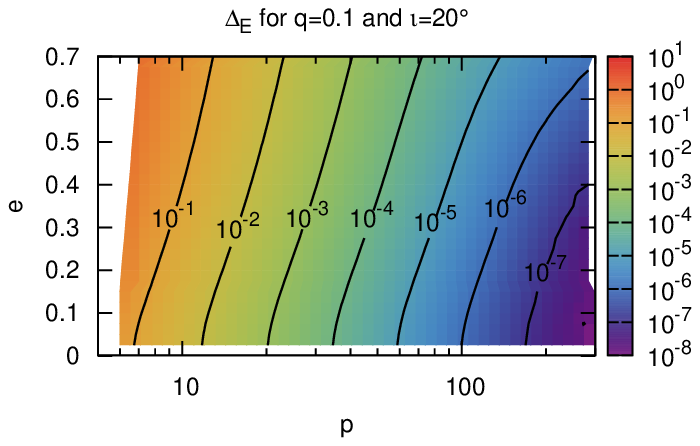}%
\includegraphics[width=51mm]{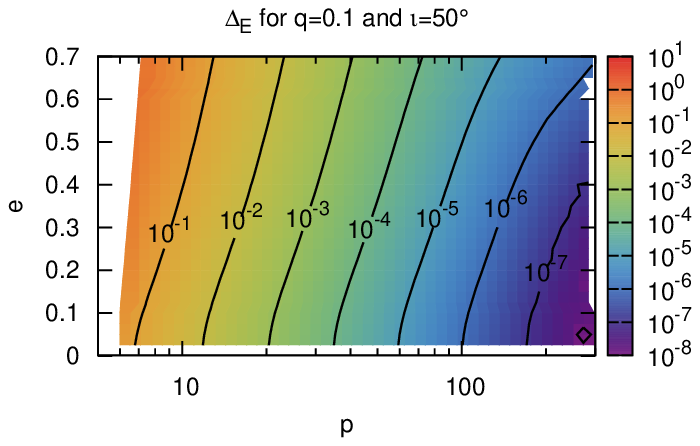}%
\includegraphics[width=51mm]{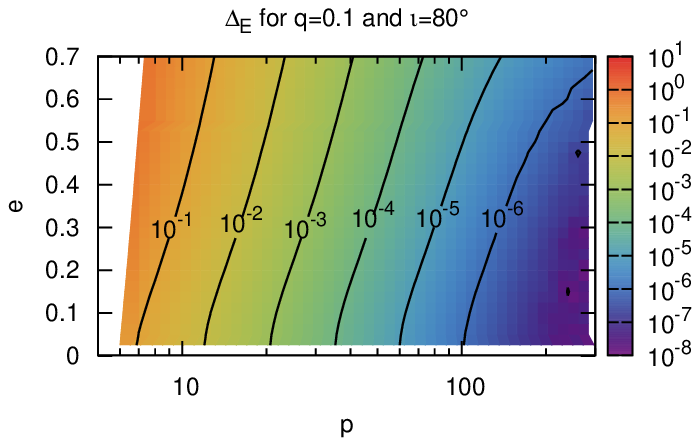}\\
\includegraphics[width=51mm]{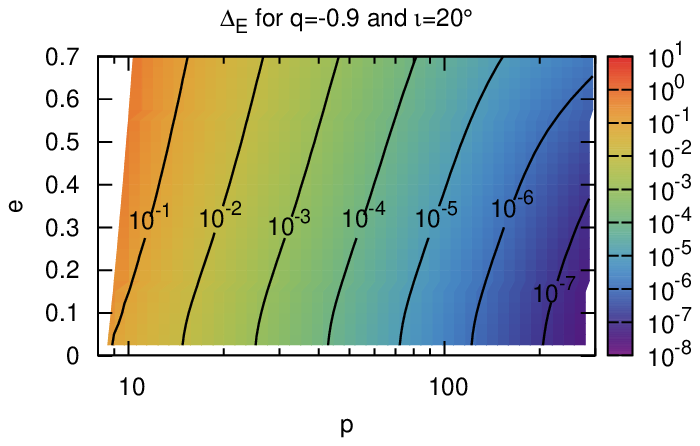}%
\includegraphics[width=51mm]{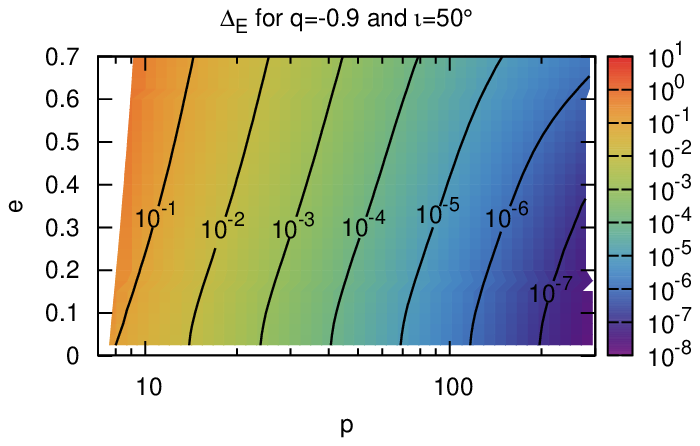}%
\includegraphics[width=51mm]{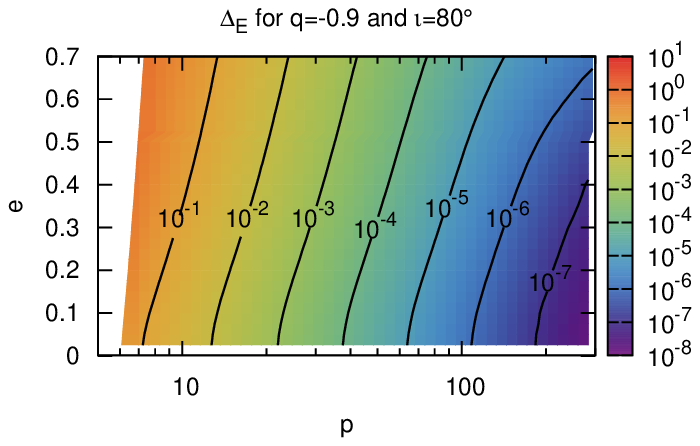}
\caption{The relative error in the 4PN $O(e^6)$ formula for
the secular change of the particle's energy, $\Delta_E$,
as a function of the semi-latus rectum $p$ and the eccentricity $e$ 
for $q=0.9, 0.5, 0.1$ and $-0.9$ (from top to bottom)
and $\iota=20^{\circ}, 50^{\circ}$ and $80^{\circ}$ (from left to right). 
We truncated the plots at $p=\max\{6, p_s(e,\iota)\}$ because the relative
errors get too large in $p<6$ to be meaningful and the orbit is not stable
for $p<p_s(e,\iota)$.
From the figures, it is easily found that the relative error becomes 
smaller (larger) for larger (smaller) $p$ and smaller (larger) $e$ with
fixed $q$ and $\iota$. 
If one requires the relative error to be less than $10^{-5}$,
the region in the semi-latus rectum $p$ and the eccentricity $e$ 
will be $p\gtrsim 50$ and $e=0.1$, 
$p\gtrsim 80$ and $e=0.4$, and $p\gtrsim 120$ and $e=0.7$ when $q=0.9$. 
It might be noticed that the relative error
does not strongly depend on the inclination angle $\iota$ 
for fixed $q$ as pointed out in Fig.~\ref{fig:dotE8H_q0.9_e0.1_0.7_inc20_80}. 
} \label{fig:dotE8H_q0.9_m0.9_inc20_80}
\end{figure}

Fig.~\ref{fig:dotE8H_q0.9_m0.9_inc20_80} shows contour plots for
$\Delta_E$ as a function of $p$ and $e$ for several sets of $(\iota, q)$.
From these plots,
one may be able to comprehend the accuracy of our PN formulae 
more easily than using Figs.~\ref{fig:dotE8H_q0.9_e0.1_0.7_inc20_80} and
\ref{fig:dotF8H_q0.9_m0.9_e0.1_0.4_0.7_inc50}. 
One will find that the relative error becomes smaller (larger) 
for larger (smaller) $p$ and smaller (larger) $e$. 
Moreover, it might be noticed that the relative error
does not strongly depend on the inclination angle $\iota$ 
for fixed $q$ as expected from Fig.~\ref{fig:dotE8H_q0.9_e0.1_0.7_inc20_80}. 
If one requires $\Delta_{E}<10^{-5}$ as an error tolerance,
one can use the contour line with the label $10^{-5}$ to find the region
of validity in the figure. For example, one will find that $\Delta_E<10^{-5}$
for $p\gtrsim 50$ and $e=0.1$, $p\gtrsim 80$ and $e=0.4$, 
and $p\gtrsim 120$ and $e=0.7$ when $q=0.9$. 

\subsection{Implementation of an exponential resummation method} 
\label{sec:exp-resum}
In order to improve the accuracy in the analytic PN formulae,
one may apply some resummation methods such as
Pad\'{e} approximation~\cite{Damour:1997ub}, 
the factorized resummation~\cite{Damour:2007xr,Damour:2007yf,Damour:2008gu}
and the exponential resummation~\cite{Isoyama:2012bx}. 
Since the exponential resummation may be the simplest one to implement
among them, we here choose to implement the exponential resummation.
We apply it to our 4PN formulae and check how the accuracy is improved.

To introduce the exponential resummation, we make use of the following identity
\begin{equation}
\left\langle\frac{dI}{dt}\right\rangle_t =
\left(\frac{dI}{dt}\right)_{\rm N}
\exp\left\{\ln\left[
\left\langle\frac{dI}{dt}\right\rangle_t
\bigg/ \left(\frac{dI}{dt}\right)_{{\rm N}}
\right]\right\}, \label{eq:exp-ln-id}
\end{equation}
where $I=\{E,L,C\}$.
The exponential resummation can be obtained by replacing the exponent
in (\ref{eq:exp-ln-id}) to the expansion with respect to $v$,
\begin{equation}
F_n^I := \ln\left[
\left\langle\frac{dI}{dt}\right\rangle_t
\bigg/ \left(\frac{dI}{dt}\right)_{{\rm N}}
\right] \bigg|_{{\rm truncated \ after \ } n{\rm th \ order \ of} \ v},
\end{equation}
where we do not perform the expansion with respect to $e$. 
Since our PN formulae for $\langle dI/dt \rangle_t$ are given at
the 4PN order, we truncate $F_n^I$ after $O(v^8)$.
Finally, the exponential resummed form is expressed as
\begin{equation}
\left\langle\frac{dI}{dt}\right\rangle_t^{\exp} =
\left(\frac{dI}{dt}\right)_N \exp F_8^I.
\end{equation}

\begin{figure}
\includegraphics[width=51mm]{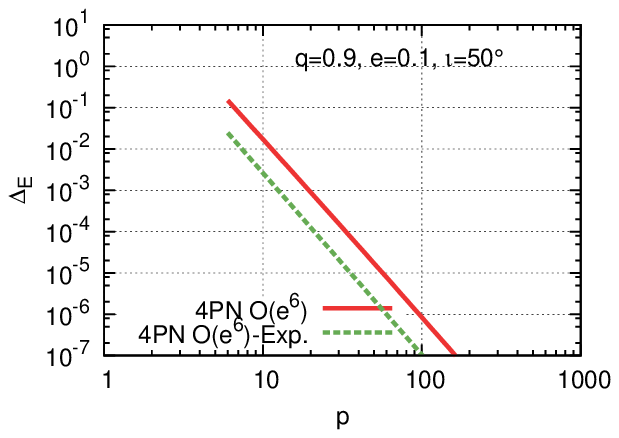}%
\includegraphics[width=51mm]{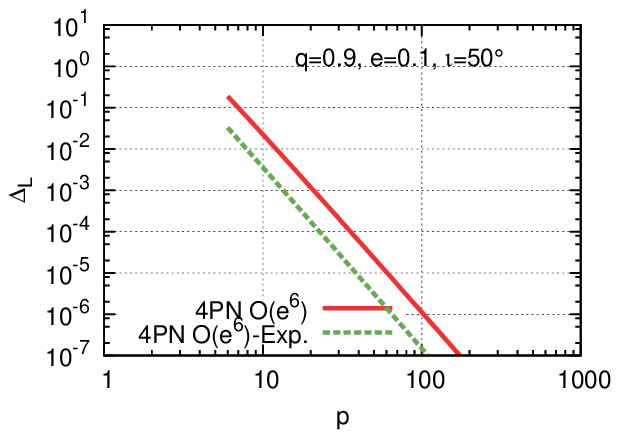}%
\includegraphics[width=51mm]{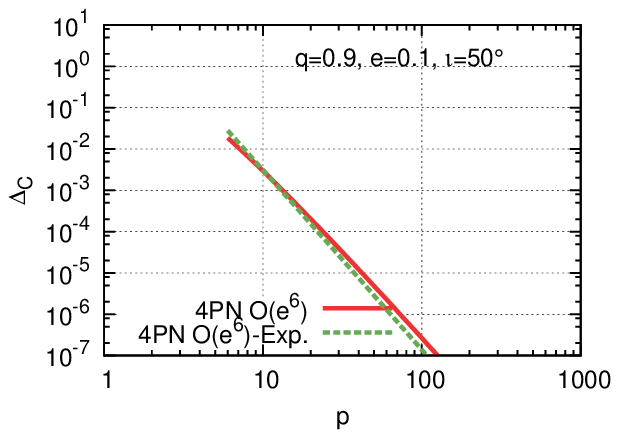}\\
\includegraphics[width=51mm]{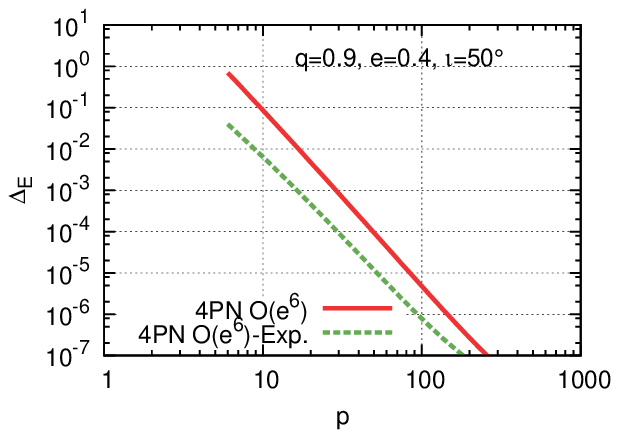}%
\includegraphics[width=51mm]{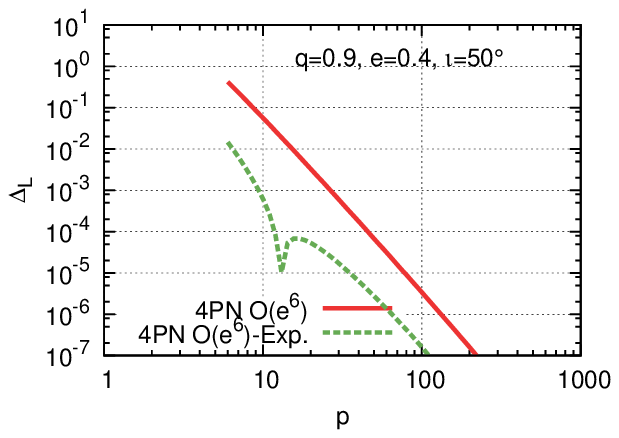}%
\includegraphics[width=51mm]{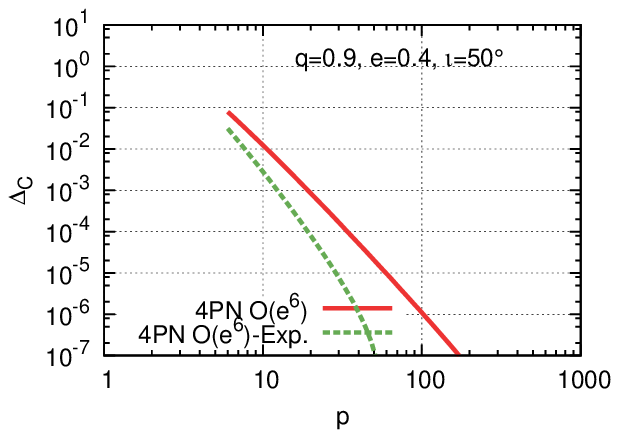}\\
\includegraphics[width=51mm]{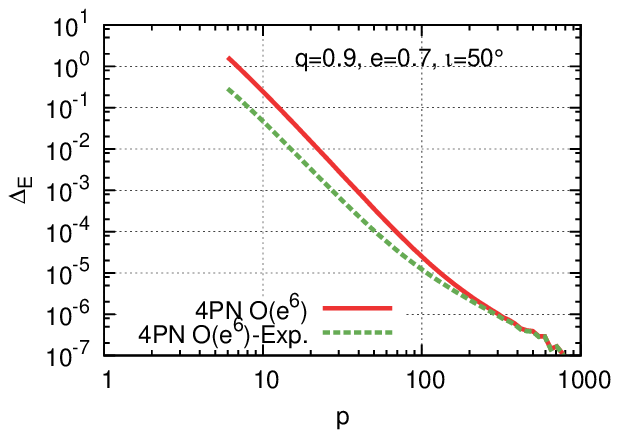}%
\includegraphics[width=51mm]{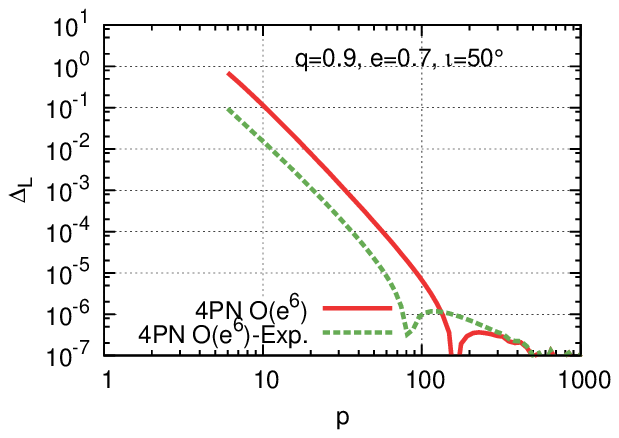}%
\includegraphics[width=51mm]{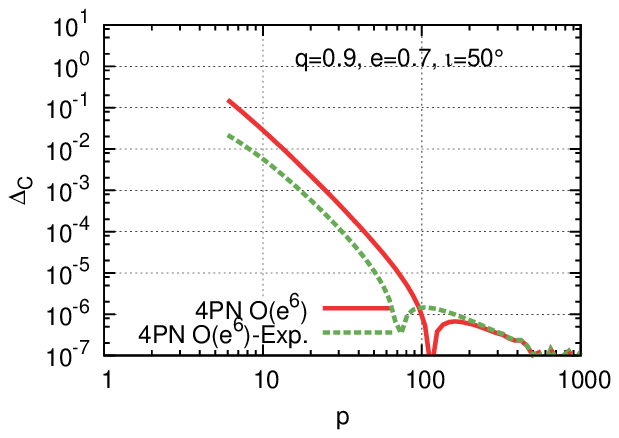}
\caption{The relative errors in the PN formulae and the exponential
resummation formulae for the secular changes of the orbital parameters,
$\{E,L,C\}$ as functions of the semi-latus rectum $p$ for $q=0.9$,
$\iota=50^{\circ}$ and $e=0.1, 0.4$ and $0.7$ (from top to bottom). 
We truncated the plots at $p=6$ because the relative errors in the
PN formulae get too large in $p<6$ to be meaningful.
Using the exponential resummation, the accuracy is improved in
most cases.
For example, the region to satisfy $\Delta_E<10^{-5}$ is improved from 
$p\gtrsim 50$ to $p\gtrsim 40$ for $e=0.1$, 
$p\gtrsim 80$ to $p\gtrsim 60$ for $e=0.4$, and 
$p\gtrsim 120$ to $p\gtrsim 100$ for $e=0.7$. 
This would suggest us to try to apply resummation methods 
to the PN formulae even in the case of general orbits. 
} \label{fig:dotF8H_q0.9_m0.9_e0.7_inc50_exp}
\end{figure}

Fig.~\ref{fig:dotF8H_q0.9_m0.9_e0.7_inc50_exp} shows the relative errors
in the exponential resummed forms of the secular changes of $E$, $L$ and
$C$, estimated by using Eq.~(\ref{eq:relative_error}).
We also show the relative
errors in the Taylor-type formulae in the same graphs for comparison.
One will find that the relative errors in the exponential resummed forms
are less than those in the Taylor-type formulae in most cases,
except for $\langle dC/dt \rangle_t$ in the case of $q=0.9$,
$(e,\iota)=(0.1, 50^\circ)$.
Using the exponential resummation when $q=0.9$ and $\iota=50^{\circ}$, 
the region to satisfy $\Delta_E<10^{-5}$ is extended to 
$p\gtrsim 40$ from $p\gtrsim 50$ for $e=0.1$, 
$p\gtrsim 60$ from $p\gtrsim 80$ for $e=0.4$, and 
$p\gtrsim 100$ from $p\gtrsim 120$ for $e=0.7$. 
This might motivate us to use the resummation method to improve the
accuracy of Taylor-type formulae even in the case of general orbits. 

\subsection{Convergence with respect to $v$ and $e$ of the analytic formulae}
\label{sec:Delta_n}
Apart from comparisons to the numerical results, 
we may also discuss the convergence property in our PN formulae 
with respect to $v$ and $e$ by investigating the contribution of
each order of $v$ and $e$ in the formulae
although this is a rough estimation. 

First we assess the PN convergence of our formulae.
For this purpose, we introduce $\Delta_n$ as
\begin{eqnarray}
\left\langle{dE \over dt}\right\rangle_t^{\rm PN}
=\left({dE\over dt}\right)_{{\rm N}} \sum_{n=0}^{8} \Delta_n,
\label{eq:Delta_n}
\end{eqnarray}
where $p=1/v^2$ and $\Delta_n$ is the $O(v^n)$ term in the
PN formula of $\langle dE/dt \rangle_t$,
{\it e.g.} $\Delta_0=1+{\frac {73}{24}}\,{e}^{2}+{\frac {37}{96}}\,{e}^{4}$,
$\Delta_1=0$ and $\Delta_2=\left( -{\frac {1247}{336}}-{\frac {9181}{672}}\,{e}^{2}+{\frac {809}{128}}\,{e}^{4}+{\frac {8609}{5376}}\,{e}^{6} \right) {v}^{2}$.
$\Delta_n$ depends on $(q,p,e,Y)$ in general although we omit the
argument for simplicity.

\begin{figure}
\includegraphics[width=51mm]{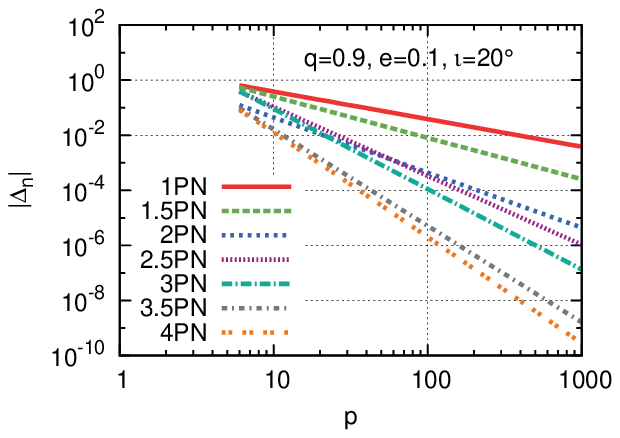}%
\includegraphics[width=51mm]{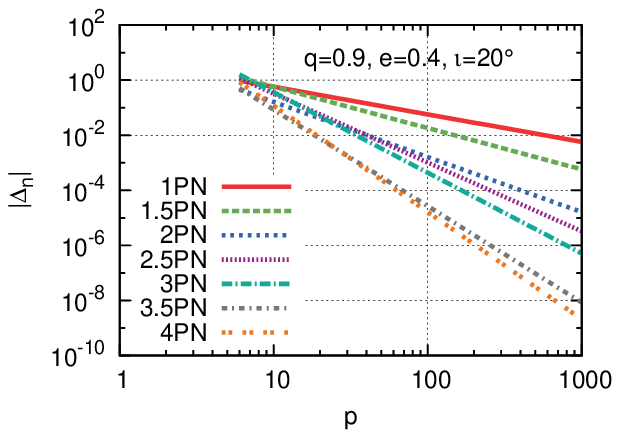}%
\includegraphics[width=51mm]{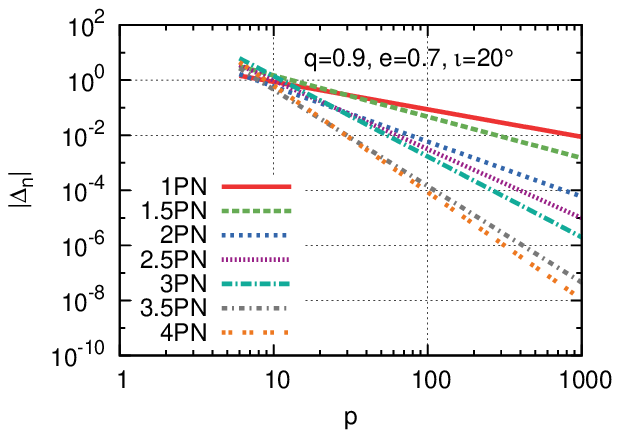}\\
\includegraphics[width=51mm]{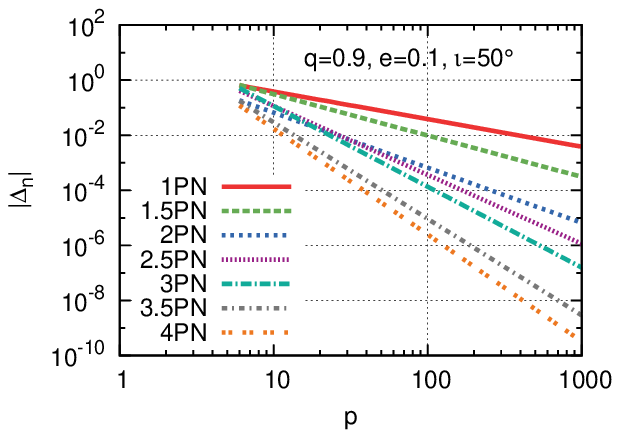}%
\includegraphics[width=51mm]{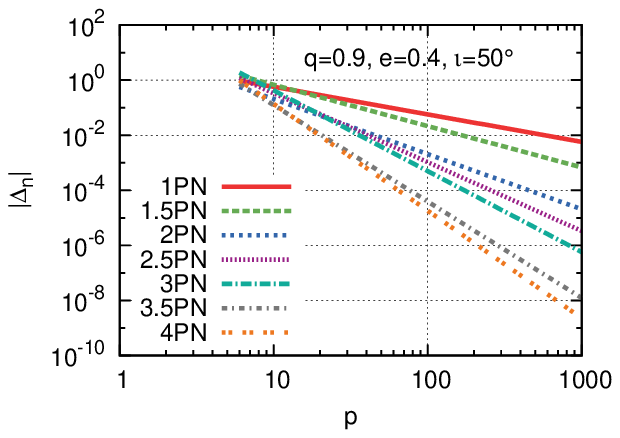}%
\includegraphics[width=51mm]{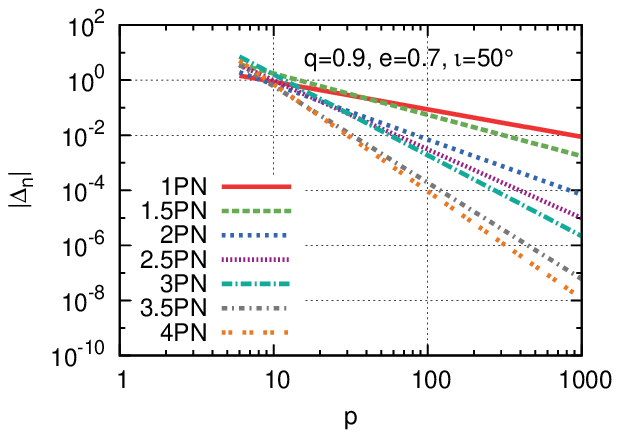}\\
\includegraphics[width=51mm]{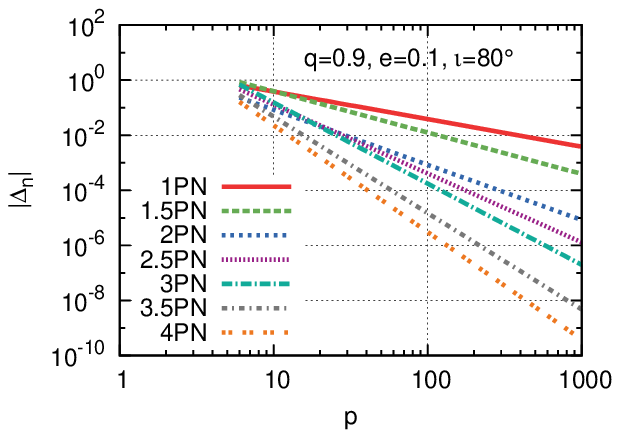}%
\includegraphics[width=51mm]{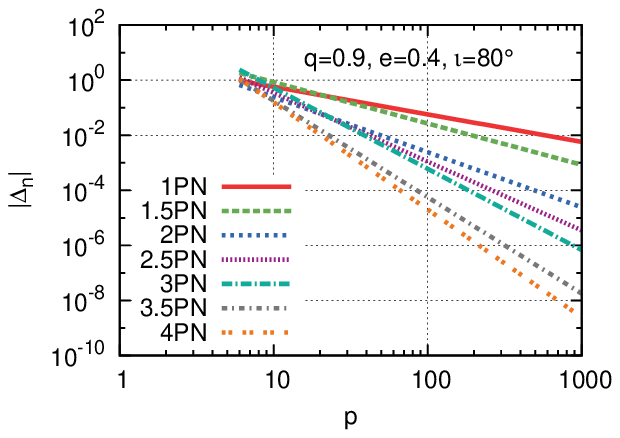}%
\includegraphics[width=51mm]{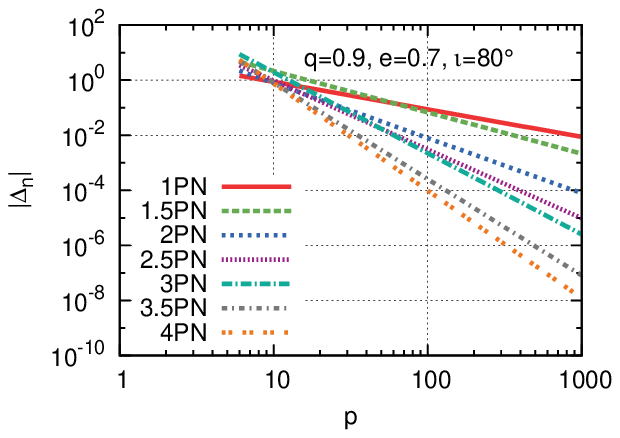}
\caption{The relative contribution of the $O(v^n)$ term in
the PN formula for $\langle dE/dt \rangle_t$,
defined in Eq.~(\ref{eq:Delta_n}).
We plot the absolute value of $\Delta_n$ as a function of the
semi-latus rectum $p$ 
for $e=0.1, 0.4$ and $0.7$ (from left to right), 
and $\iota=20^\circ, 50^\circ$ and $80^\circ$ (from top to bottom)
when $q=0.9$. 
We truncated the plots at $p=6$ because the relative contributions
get too large in $p<6$ to be meaningful.
It is expected that $|\Delta_{n+1}|<|\Delta_{n}|$ for moderately
large $n$ if the PN formula converges.
As shown in Sec.~\ref{sec:compare}, $\Delta_n$ does not strongly depend
on $\iota$ for a fixed $e$ although $\Delta_n$ strongly depends on $e$. 
In fact, the convergence seems worse the orbital eccentricity 
becomes larger. This tendency is clear for small $p$,
{\it e.g.} $p\lesssim 10$.
} \label{fig:dotE_vn_q0.9_e0.1_0.9_Y1_0}
\end{figure}

Since $\Delta_n$ shows the relative importance of the $O(v^n)$ term
in the PN formulae, it can be used to investigate the convergence
property with respect to $v$:
it is expected that $|\Delta_{n+1}|<|\Delta_{n}|$ for moderately large $n$
if the PN formula converges.
In Fig.~\ref{fig:dotE_vn_q0.9_e0.1_0.9_Y1_0}, we plot the relative
contribution of each order, $\Delta_n$, as a function of $p$ for
several sets of $(e,\iota)$ and $q=0.9$.
From this figure, one may find that $\Delta_n$ does not strongly
depend on the inclination angle, $\iota$, as shown in Sec.~\ref{sec:compare},
while it strongly depends on $e$.
The convergence gets worse when the orbital eccentricity becomes larger.
This tendency is particularly evident in the small-$p$ region.
Fixing the value of $p$, the orbit with larger $e$ passes by
closer to the central black hole and will be affected by the stronger
gravitational field. Hence the PN convergence is expected to be worse
when the eccentricity becomes larger.

Next, in order to investigate the convergence of the expansion with
respect to the orbital eccentricity in the PN formula, we introduce $A_n$
as 
\begin{eqnarray}
\left\langle{dE \over dt}\right\rangle_t^{\rm PN}
=\left({dE\over dt}\right)_{{\rm N}}\,[A_0\,e^0+A_2\,e^2+A_4\,e^4+A_6\,e^6],
\label{eq:A_n}
\end{eqnarray}
where the term $A_0$ coincides with the energy flux for circular orbits
and $A_n=0$ when $n$ is odd. 

\begin{figure}
\includegraphics[width=51mm]{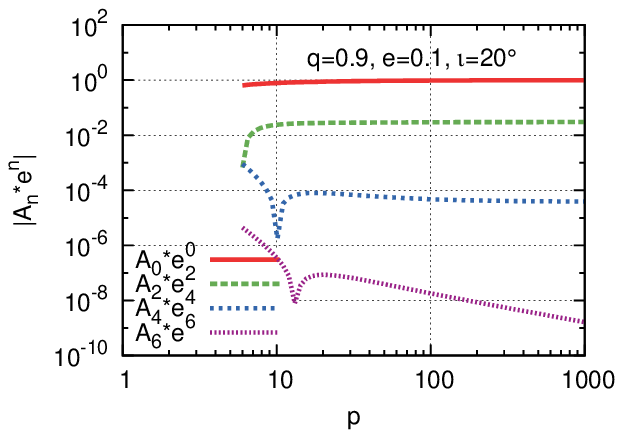}%
\includegraphics[width=51mm]{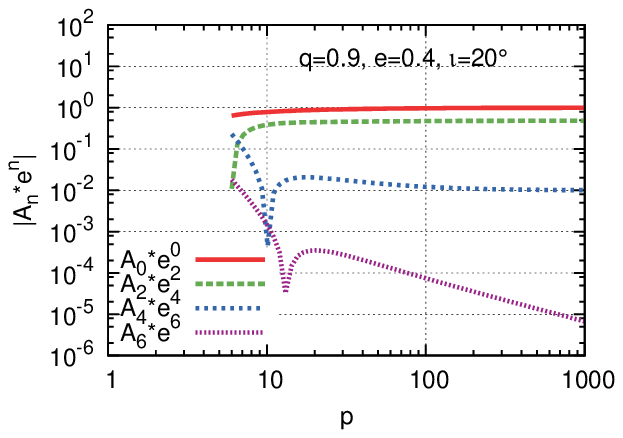}%
\includegraphics[width=51mm]{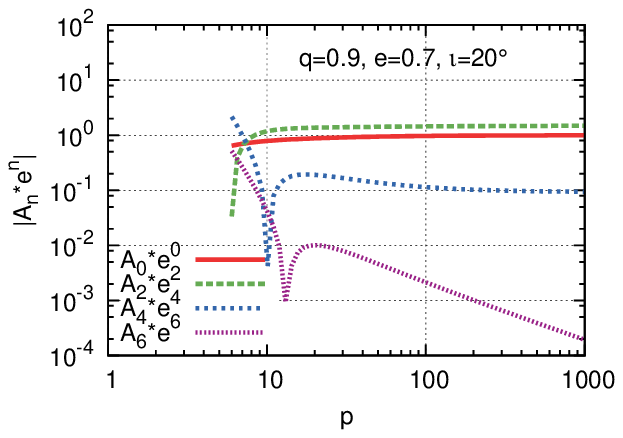}\\
\includegraphics[width=51mm]{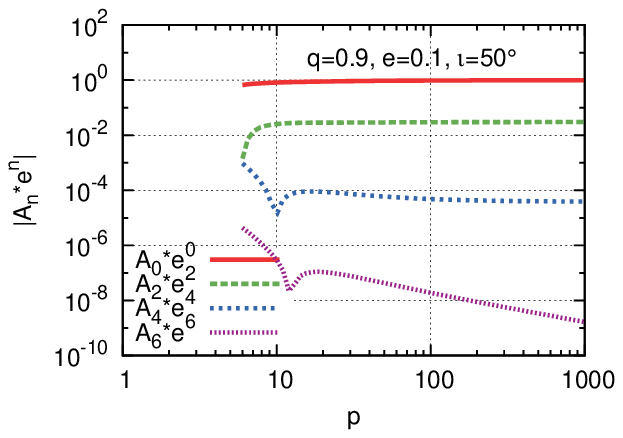}%
\includegraphics[width=51mm]{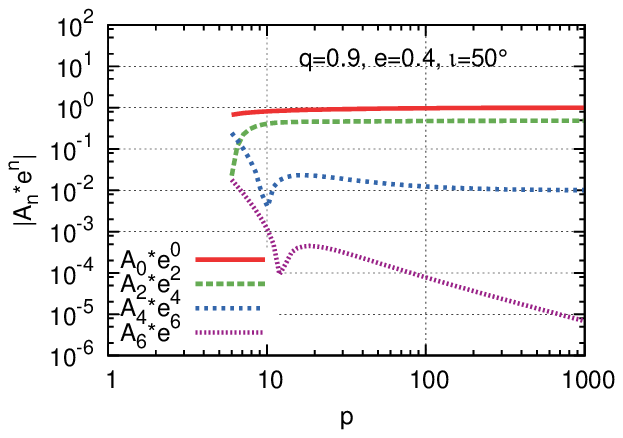}%
\includegraphics[width=51mm]{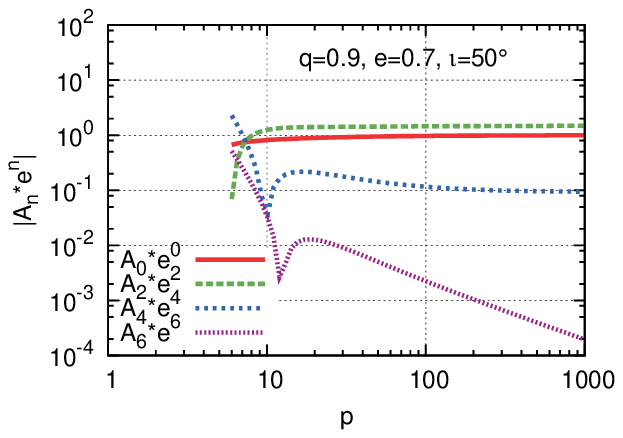}\\
\includegraphics[width=51mm]{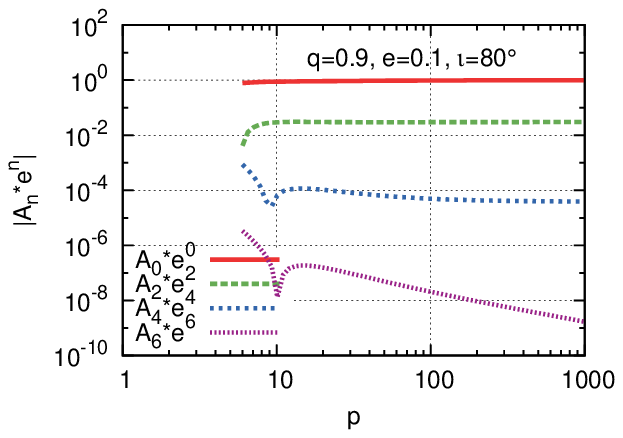}%
\includegraphics[width=51mm]{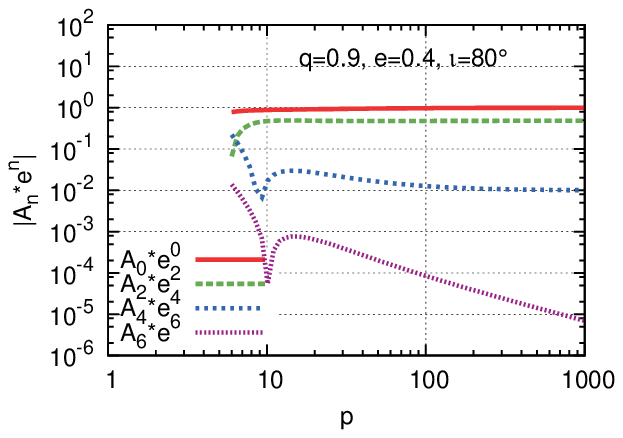}%
\includegraphics[width=51mm]{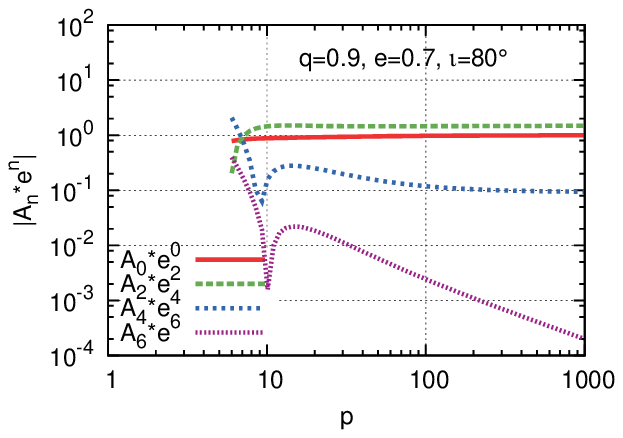}
\caption{The relative contribution of the $O(e^{n})$ term in the
PN formula for $\langle dE/dt \rangle_t$,
defined in Eq.~(\ref{eq:A_n}).
We plot the absolute value of $A_{n} e^{n}$ as a function of the
semi-latus rectum $p$ for 
$\iota = 20^\circ, 50^\circ$ and $80^\circ$ (from left to right) 
when $q=0.9$. 
We truncated the plots at $p=6$ because the relative contributions get
too large in $p<6$ to be meaningful.
It is expected that $|A_{2n+2}e^{2n+2}|<|A_{2n}e^{2n}|$ for moderately
large $n$ if the series with respect to $e$ converges.
This condition is satisfied in most cases shown in this figure.
The convergence becomes slower when the eccentricity is
larger. Especially, the convergence for $p\lesssim 10$ is quite
bad in $e=0.7$ case. 
We also note that $A_n$ does not strongly depend on $\iota$ 
for a fixed $q$ as mentioned in Sec.~\ref{sec:compare}. 
} \label{fig:dotE_en_q0.9_Y1_0}
\end{figure}

One may ask whether the condition, $|A_{2n+2}e^{2n+2}|<|A_{2n}e^{2n}|$, is
satisfied for moderately large integer $n$ if the series converges.
From Fig.~\ref{fig:dotE_en_q0.9_Y1_0},
it is found that the condition is satisfied in most cases.
As expected, the convergence becomes slower when the eccentricity is
larger. Especially, the convergence gets worse when $p\lesssim 10$
in $e=0.7$ case. 
The calculation of the higher PN corrections will be
necessary to improve the bad convergence for small $p$.
We also note that $A_n$ does not strongly depend on $\iota$ 
for a fixed $q$ as in Sec.~\ref{sec:compare}. 

\section{Summary} \label{sec:summary}
We have derived the secular changes of the orbital parameters,
the energy, azimuthal angular momentum, and Carter parameter
of a point particle orbiting a Kerr black hole,
by using the post-Newtonian approximation in the first order black
hole perturbation theory. 
We have extended the previous work~\cite{Ganz:2007rf}, 
which derived formulae up to the 2.5PN order with
the second order correction with respect to the eccentricity,
to the 4PN order with
the sixth order correction with respect to the eccentricity. 
We have also included the contribution due to the black hole absorption,
which has not been included in \cite{Ganz:2007rf}.
As shown in the case of equatorial, circular orbits
~\cite{Poisson:1994yf,Tagoshi:1997jy}, 
we have found that the secular changes of the three orbital parameters
due to the absorption, appear at the 2.5PN (4PN) from the leading order 
in the Kerr (Schwarzschild) case, and that the 2.5PN and 3.5PN
contributions of the absorption to the secular change of the particle's
energy can be positive for $q>0$, which implies that a superradiance can
be realized in the Kerr case.
We have also found that the superradiant contributions in the secular
change of the energy get smaller when the inclination angle becomes larger
and they vanishes for polar ($Y=0$) orbits. 
This means that the superradiant scattering may be 
suppressed for inclined orbits~\cite{Hughes:2001jr}. 

To investigate the accuracy in our 4PN formulae,
we have compared the formulae to high-precision numerical 
results~\cite{Fujita:2009us} in Sec.~\ref{sec:compare}. 
We have found that the accuracy gets worse when the orbital velocity and
the orbital eccentricity become larger, as expected.
If the relative error in the 4PN $O(e^6)$ formula for the secular change
of the energy is required to be less than $10^{-5}$, the parameter region
to satisfy it might be $p\gtrsim 50$ for $e=0.1$, 
$p\gtrsim 80$ for $e=0.4$, and $p\gtrsim 120$ for $e=0.7$ when $q=0.9$. 
The region does not strongly depend on the orbital inclination angle. 
From Fig.~\ref{fig:dotE8H_q0.9_e0.1_0.7_inc20_80},
one can clearly find the improvement of the accuracy in our PN formulae 
from the previous work at the 2.5PN order and the second order correction
in the orbital eccentricity~\cite{Ganz:2007rf} whose relative error is
larger than $10^{-2}$ for $p\gtrsim 100$ when $e\gtrsim 0.4$ 
since, in this region, the error due to the truncation of the expansion
with respect to the orbital eccentricity is larger than
the one of the PN expansion.

One may improve the accuracy of our PN formulae 
by using resummation methods. In this paper, we have applied 
the exponential resummation~\cite{Isoyama:2012bx} to our 4PN formulae
and confirmed that the resummation method improves the accuracy in
most cases investigated here.
For example, we found that the region in which the relative errors are
less than $10^{-5}$ can be extended from 
$p\gtrsim 50$ to $p\gtrsim 40$ for $e=0.1$, 
$p\gtrsim 80$ to $p\gtrsim 60$ for $e=0.4$, and 
$p\gtrsim 120$ to $p\gtrsim 100$ for $e=0.7$.

We also investigate the convergence properties of the PN expansion and the
expansion with respect to the orbital eccentricity, respectively.
Both convergences get worse when the semi-latus rectum is smaller; in
other words, the gravitational field becomes stronger.
This tendency gets clearer in the case of large eccentricity, in which
the particle passes by closer to the central black hole.

In order to improve the accuracy and convergence of the 4PN $O(e^6)$ formulae
near the central black hole and to obtain the physical information of 
the source in the strong-field region, it is necessary to derive 
the higher order corrections of the PN expansion and the expansion with
respect to the eccentricity.
It may be possible to avoid the expansion with respect to the eccentricity
and to derive the PN formulae applicable to arbitrary eccentricity.
So far the PN formulae of the rate of the energy loss without performing
the expansion with respect to the eccentricity had been derived for equatorial
orbits in~\cite{Galtsov:1980ef,Shibata:1994xk,Arun:2007sg,Arun:2009mc}. 
The extension of these results to the case of inclined orbits is challenging:
we can obtain analytic expressions for general bound geodesic orbits in Kerr
spacetime without performing the expansion with respect to the eccentricity nor
the inclination by using results in Ref.~\cite{Fujita:2009bp}, while we need
to reformulate the source term of the Teukolsky equation and the derivation of
the partial waves constructed form the source term.
We would like to leave it to the future work.

\section*{Acknowledgment}
We would like to thank Takahiro Tanaka and Hiroyuki Nakano for useful
discussions and comments. We are also grateful to Theoretical Astrophysics
Group in Kyoto University for hospitality during the intermediate stage
of completing this paper. 
NS acknowledges the support of the Grand-in-Aid for Scientific Research 
(No. 25800154). RF's work was supported by 
the European Union's FP7 ERC Starting Grant ``The dynamics of black holes:
testing the limits of Einstein's theory'' grant agreement no. DyBHo--256667. 
Some numerical computations were performed at the cluster 
``Baltasar-Sete-S\'{o}is'' in CENTRA/IST. 
Some analytic calculations were carried out on HA8000/RS440 at 
Yukawa Institute for Theoretical Physics in Kyoto University. 

\appendix
\section{PN formulae for the orbital parameters and fundamental
frequencies} \label{App:variable_exp}
In this section, we present the PN formulae of the orbital parameters,
$\{\hat{E}, \hat{L}, \hat{C}\}$, and the fundamental frequencies,
$\{\Omega_t, \Omega_r, \Omega_\theta, \Omega_\phi\}$.
Here we show the formulae up to the 3PN $O(e^6)$ order to save space
although it is possible to calculate them to the higher order.
The higher order results will be publicly available online \cite{BHPC}.
\begin{eqnarray}
\hat{E} &=& 1+ \left\{ -\frac{1}{2}+\frac{1}{2}\,{e}^{2} \right\} {v}^{2}
+ \left\{ \frac{3}{8} -\frac{3}{4}\,{e}^{2}+\frac{3}{8}\,{e}^{4}\right\} {v}^{4}
+ \left\{ -Yq +2\,Yq\,{e}^{2}-Yq\,{e}^{4}\right\} {v}^{5}
\nonumber \\ &&
+ \left\{ {\frac {27}{16}}+\frac{1}{2}\,{Y}^{2}{q}^{2}+ \left( -{\frac {49}{16}}-{Y}^{2}{q}^{2} \right) {e}^{2}+ \left( {\frac {17}{16}}+\frac{1}{2}\,{Y}^{2}{q}^{2} \right) {e}^{4}+{\frac {5}{16}}\,{e}^{6} \right\} {v}^{6}, \\
\hat{L} &=& p\,v\biggl[
Y+ \left\{ \frac{3}{2}\,Y+\frac{1}{2}\,Y{e}^{2} \right\} {v}^{2}
+ \left\{ -3\,{Y}^{2}q-q{Y}^{2}{e}^{2} \right\} {v}^{3}
\nonumber \\ && \hspace{0.5cm}
+ \left\{ {\frac {27}{8}}\,Y+{Y}^{3}{q}^{2}+ \left( \frac{9}{4}\,Y+{Y}^{3}{q}^{2} \right) {e}^{2}+\frac{3}{8}\,Y{e}^{4} \right\} {v}^{4}
\nonumber \\ && \hspace{0.5cm}
+ \left\{ -\frac{15}{2}\,{Y}^{2}q-7\,q{Y}^{2}{e}^{2}-\frac{3}{2}\,{Y}^{2}q{e}^{4
} \right\} {v}^{5}
\nonumber \\ && \hspace{0.5cm}
+ \biggl\{ {\frac {135}{16}}\,Y-4\,Y{q}^{2}+9\,{Y}^{3}{
q}^{2}+ \left( {\frac {135}{16}}\,Y-4\,Y{q}^{2}+9\,{Y}^{3}{q}^{2}
 \right) {e}^{2} \hspace{0.5cm}
\nonumber \\ && \hspace{1.0cm}
+ \left( {\frac {45}{16}}\,Y+2\,{Y}^{3}{q}^{2}
 \right) {e}^{4}+{\frac {5}{16}}\,Y{e}^{6} \biggr\} {v}^{6} \biggr], \\
\hat{C} &=& \left\{\frac{1}{Y^2}-1\right\}\,L^2, \nonumber \\ 
&=&
p^2 v^2 (1-Y^2) \Bigl[
 1 + (3+e^2) v^2 - 2 q Y \left( 3+e^2 \right) v^3
\nonumber \\ &&
 + \left\{ 9+2\,{Y}^{2}{q}^{2}+ \left( 6+2\,{Y}^{2}{q}^{2} \right) {e}^{2}
 +{e}^{4} \right\} v^4
 - 4\,qY \left( 2+{e}^{2} \right)  \left( 3+{e}^{2} \right) v^5
\nonumber \\ &&
 + \bigl\{ 27-8\,{q}^{2}+30\,{Y}^{2}{q}^{2}+ \left( 27+28\,{Y}^{2}{q}^{2}
    -8\,{q}^{2} \right) {e}^{2}
 + \left( 9+6\,{Y}^{2}{q}^{2} \right) {e}^{4}
    +{e}^{6} \bigr\} v^6 \Bigr], \\
\Omega_t &=& p^2 \biggl[
1+\frac{3}{2}\,{e}^{2}+{\frac {15}{8}}\,{e}^{4}+{\frac {35}{16}}\,{e}^{6}+
 \left\{ \frac{3}{2}-\frac{1}{4}\,{e}^{2}-{\frac {15}{16}}\,{e}^{4}-{\frac {45}{32}}\,{e}^{6} \right\} {v}^{2}
\nonumber \\ && \hspace{0.5cm}
+ \left\{ 2\,Yq\,{e}^{2}+3\,Yq\,{e}^{4}+{\frac {15}{4}}\,Yq\,{e}^{6} 
\right\} {v}^{3}
\nonumber \\ && \hspace{0.5cm}
+\biggl\{ {\frac {27}{8}}-\frac{1}{2}\,{Y}^{2}{q}^{2}
+\frac{1}{2}\,{q}^{2}+ \left( -{\frac {99}{16}}+{q}^{2}-2\,{Y}^{2}{q}^{2}
 \right) {e}^{2}
\nonumber \\ && \hspace{1.0cm}
+ \left( -{\frac {567}{64}}+{\frac {21}{16}}\,{q}^{2}-
{\frac {45}{16}}\,{Y}^{2}{q}^{2} \right) {e}^{4}+ \left( -{\frac {1371
}{128}}+{\frac {25}{16}}\,{q}^{2}-{\frac {55}{16}}\,{Y}^{2}{q}^{2}
 \right) {e}^{6} \biggr\} {v}^{4}
\nonumber \\ && \hspace{0.5cm}
+ \left\{ -3\,Yq+{\frac {43}{2}}\,Yq\,{e}^{2}+{\frac {231}{8}}\,Yq\,{e}^{4}+{\frac {555}{16}}\,Yq\,{e}^{6} \right\} {v}^{5}
\nonumber \\ && \hspace{0.5cm}
+ \biggl\{ {\frac {135}{16}}-\frac{1}{4}\,{q}^{2}+\frac{3}{4}\,{Y}^{2}{q}^{2}+
 \left( -{\frac {1233}{32}}+{\frac {47}{4}}\,{q}^{2}-{\frac {75}{2}}\,
{Y}^{2}{q}^{2} \right) {e}^{2}
\nonumber \\ &&\hspace{1.0cm}
+ \left( -{\frac {6567}{128}}+{\frac {
499}{32}}\,{q}^{2}-{\frac {1577}{32}}\,{Y}^{2}{q}^{2} \right) {e}^{4}
\nonumber \\ &&\hspace{1.0cm}
+ \left( -{\frac {15565}{256}}+{\frac {75}{4}}\,{q}^{2}-{\frac {1887}{
32}}\,{Y}^{2}{q}^{2} \right) {e}^{6} \biggr\} {v}^{6} \biggr], \\
\Omega_r &=& p\,v\biggl[
1+ \left\{ -\frac{3}{2}+\frac{1}{2}\,{e}^{2} \right\} {v}^{2}
+ \left\{ 3\,Yq-Yq\,{e}^{2} \right\} {v}^{3}
\nonumber \\ && \hspace{0.5cm}
+ \left\{ -{\frac {45}{8}}+\frac{1}{2}\,{q}^{2}-2\,{Y}^{2}{q}^{2}
+ \left( \frac{1}{4}\,{q}^{2}+\frac{1}{4}\,{Y}^{2}{q}^{2} \right) {e}^{2}+
\frac{3}{8}\,{e}^{4} \right\} {v}^{4}
\nonumber \\ && \hspace{0.5cm}
+ \left\{ {\frac {33}{2}}\,Yq +2\,Yq\,{e}^{2}-\frac{3}{2}\,Yq\,{e}^{4}\right\} {v}^{5}
\nonumber \\ && \hspace{0.5cm}
+ \biggl\{ -{\frac {351}{16}} -{\frac {51}{2}}\,{Y}^{2}{q}^{2}
+{\frac {33}{4}}\,{q}^{2}+ \left( -{\frac {135}{16}}+{\frac {
7}{8}}\,{q}^{2}-{\frac {39}{8}}\,{Y}^{2}{q}^{2} \right) {e}^{2}
\nonumber \\ && \hspace{1.0cm}
+ \left( {\frac {21}{16}}+\frac{1}{8}\,{q}^{2}+{\frac {13}{8}}\,{Y}^{2}{q}^{2}
 \right) {e}^{4}+{\frac {5}{16}}\,{e}^{6} \biggr\} {v}^{6} \biggr], \\
\Omega_\theta &=& p\,v\biggl[
1+ \left\{ \frac{3}{2}+\frac{1}{2}\,{e}^{2}\right\} {v}^{2}- \left\{ 3\,Yq+Yq\,{e}^{2} \right\} {v}^{3}
\nonumber \\ && \hspace{0.5cm}
+ \left\{ {\frac {27}{8}}+\frac{7}{4}\,{Y}^{2}{q}^{2}-\frac{1}{4}\,{q}^{2}+ \left( \frac{9}{4}+\frac{1}{4}\,{q}^{2}+\frac{1}{4}\,{Y}^{2}{q}^{2} \right) {e}^{2}+\frac{3}{8}\,{e}^{4} \right\} {v}^{4}
\nonumber \\ && \hspace{0.5cm}
- \left\{ \frac{15}{2}\,Yq+7\,Yq\,{e}^{2}+\frac{3}{2}\,Yq\,{e}^{4}
 \right\} {v}^{5}
\nonumber \\ && \hspace{0.5cm}
+ \biggl\{ {\frac {135}{16}}+{\frac {57}{8}}\,{Y}^{2}{q}^{2}
-{\frac {27}{8}}\,{q}^{2}+ \left( {\frac {135}{16}}-{\frac {19}{4}
}\,{q}^{2}+{\frac {45}{4}}\,{Y}^{2}{q}^{2} \right) {e}^{2}
\nonumber \\ && \hspace{1.0cm}
+ \left( {
\frac {45}{16}}+\frac{1}{8}\,{q}^{2}+{\frac {13}{8}}\,{Y}^{2}{q}^{2} \right) 
{e}^{4}+{\frac {5}{16}}\,{e}^{6} \biggr\} {v}^{6} \biggr], \\
\Omega_\varphi &=& p\,v\biggl[
1+ \left\{ \frac{3}{2}+\frac{1}{2}\,{e}^{2} \right\} {v}^{2}
+ \left\{ 2\,q-3\,Yq -Yq\,{e}^{2}\right\} {v}^{3}
\nonumber \\ && \hspace{0.5cm}
+ \left\{ -\frac{3}{2}\,Y{q}^{2}+\frac{7}{4}\,{Y}^{2}{q}^{2}-\frac{1}{4}\,{q}^{2}+{\frac {27}{8}}+ \left( \frac{9}{4}+\frac{1}{4}\,{q}^{2}+\frac{1}{4}\,{Y}^{2}{q}^{2} \right) {e}^{2}+\frac{3}{8}\,{e}^{4} \right\} {v}^{4}
\nonumber \\ && \hspace{0.5cm}
+ \left\{ 3\,q-\frac{15}{2}\,Yq+ \left( 4\,q-7\,Yq \right) {e}^{2}-\frac{3}{2}\,Yq\,{e}^{4} \right\} {v}^{5}
\nonumber \\ && \hspace{0.5cm}
+ \biggl\{ -\frac{9}{4}\,Y{q}^{2}+{\frac {57}{8}}\,{Y}^{2}{q}^{2}+{\frac {135}{16
}}-{\frac {27}{8}}\,{q}^{2}+ \left( {\frac {135}{16}}-{\frac {19}{4}}
\,{q}^{2}-{\frac {35}{4}}\,Y{q}^{2}+{\frac {45}{4}}\,{Y}^{2}{q}^{2}
 \right) {e}^{2} 
\nonumber \\ && \hspace{1.0cm}
+ \left( {\frac {45}{16}}+\frac{1}{8}\,{q}^{2}+{\frac {13}{8}}
\,{Y}^{2}{q}^{2} \right) {e}^{4}+{\frac {5}{16}}\,{e}^{6} \biggr\} {v}^{6}
 \biggr].
\end{eqnarray}

\section{Fourier coefficients of bound orbits} \label{App:coeff_orbit}
Here we show the PN formulae of the Fourier coefficients in
Eqs.~(\ref{eq:r-fourier}), (\ref{eq:theta-fourier}),
(\ref{eq:t-fourier}) and (\ref{eq:phi-fourier}) up to the 3PN $O(e^6)$ order.
The 4PN $O(e^6)$ results obtained in this work will be available
online \cite{BHPC}.

In this work, we follow the same procedure as in \cite{Ganz:2007rf}
to derive the amplitudes of the partial waves,
$Z_\Lambda^{{\rm H},\infty}$ in (\ref{eq:partial-wave}).
In the formal expression, the dependence of $\varphi^{(\theta)}$ appears
in the form of the combination as $X\equiv\sin\theta
{\rm e}^{i\varphi^{(\theta)}}$, which can be expressed in the Fourier
series as
\begin{equation}
X = p \sum_{n_\theta=0}^\infty \left[
X_{n_\theta}^\Re \cos n_\theta \Omega_\theta \lambda
+ i X_{n_\theta}^\Im \sin n_\theta \Omega_\theta \lambda
\right],
\end{equation}
Therefore we show the Fourier coefficients of $X$ instead of
$\varphi^{(\theta)}$.

\subsection{Radial component}
\begin{eqnarray}
\alpha_0 &=& 1
+ e^2 \biggl\{
\frac{1}{2}
- \frac{1}{2}{v}^{2}
+ qY{v}^{3}
+ \left( -3 + \left( \frac{1}{2}
 - {Y}^{2} \right) {q}^{2} \right) {v}^{4}
\nonumber \\ &&
+ 10\,qY{v}^{5}
+ \left( - 18
 + \left( \frac{11}{2} - 18\,{Y}^{2} \right) {q}^{2}
 \right) {v}^{6}
\biggr\}
\nonumber \\ &&
+ e^4 \biggl\{
 \frac{3}{8}
 - \frac{3}{8}\,{v}^{2}
 + \frac{3}{4}\,qY{v}^{3}
 + \left( -{\frac {33}{16}}
   + \left( {\frac {5}{16}}-{\frac {11}{16}}\,{Y}^{2} \right)
    {q}^{2} \right) {v}^{4}
\nonumber \\ &&
 + {\frac {27}{4}}\,qY{v}^{5}
 + \left( -{\frac {189}{16}} + \left( -{\frac {185}{16}}\,{Y}^{2}
   + {\frac {61}{16}} \right) {q}^{2} \right) {v}^{6}
\biggr\}
\nonumber \\ &&
+ e^6 \biggl\{
 {\frac {5}{16}}
 - {\frac {5}{16}}\,{v}^{2}
 + \frac{5}{8}\,qY{v}^{3}
 + \left( -{\frac {27}{16}}
         + \left( \frac{1}{4}
           -{\frac {9}{16}}\,{Y}^{2} \right) {q}^{2}
   \right) {v}^{4}
\nonumber \\ &&
 + \frac{11}{2}\,qY{v}^{5}
 + \left( -\frac{19}{2}
    + \left( -{\frac {75}{8}}\,{Y}^{2}+{\frac {49}{16}}
    \right) {q}^{2} \right) {v}^{6}
\biggr\}, \\
\alpha_1 &=& e
+ e^3 \biggl\{
 \frac{3}{4}
 - \frac{1}{2}\,{v}^{2}
 + Yq{v}^{3}
 + \left( -{\frac {51}{16}}+ \left( {\frac {7}{16}}
    -{\frac {15}{16}}\,{Y}^{2} \right) {q}^{2} \right) {v}^{4}
\nonumber \\ &&
 + {\frac {43}{4}}\,Yq{v}^{5}
 + \left( -{\frac {81}{4}}+ \left( -{\frac {153}{8}}\,{Y}^{2}
    + \frac{11}{2} \right) {q}^{2} \right) {v}^{6}
\biggr\}
\nonumber \\ &&
+ e^5 \biggl\{
 \frac{5}{8}
 - \frac{1}{2}\,{v}^{2}
 + Yq{v}^{3}
 + \left( -{\frac {93}{32}}+ \left( {\frac {13}{32}}
    -{\frac {29}{32}}\,{Y}^{2} \right) {q}^{2} \right) {v}^{4}
\nonumber \\ &&
 + {\frac {77}{8}}\,Yq{v}^{5}
 + \left( -{\frac {277}{16}}+ \left( -{\frac {133}{8}}\,{Y}^{2}
    + {\frac {83}{16}} \right) {q}^{2} \right) {v}^{6}
\biggr\}, \\
\alpha_2 &=& 
e^2 \biggl\{
 \frac{1}{2}
 + \frac{1}{2}\,{v}^{2}
 - Yq{v}^{3}
 + \left( 3+ \left( {Y}^{2} - \frac{1}{2} \right) {q}^{2}
   \right) {v}^{4}
\nonumber \\ &&
 - 10\,Yq{v}^{5}
 + \left( 18+ \left( 18\,{Y}^{2} - \frac{11}{2} \right) {q}^{2}
   \right) {v}^{6}
\biggr\}
\nonumber \\ &&
+ e^4 \biggl\{
 \frac{1}{2}
 - \frac{1}{2}\,{v}^{4}
 + 2\,Yq{v}^{5}
 + \left( -\frac{11}{2}+ \left( -4\,{Y}^{2} + \frac{1}{2}
      \right) {q}^{2} \right) {v}^{6}
\biggr\}
\nonumber \\ &&
+ e^6 \biggl\{
 {\frac {15}{32}}
 - {\frac {5}{32}}\,{v}^{2}
 + {\frac {5}{16}}\,Yq{v}^{3}
 + \left( -{\frac {39}{32}}
     + \left( \frac{1}{8}-{\frac {9}{32}}\,{Y}^{2}
     \right) {q}^{2} \right) {v}^{4}
\nonumber \\ &&
 + {\frac {17}{4}}\,Yq{v}^{5}
 + \left( -{\frac {279}{32}}
     + \left( -{\frac {243}{32}}\,{Y}^{2}+2 \right) {q}^{2}
    \right) {v}^{6}
\biggr\}, \\
\alpha_3 &=&
e^3 \biggl\{
 \frac{1}{4}
 + \frac{{v}^{2}}{2}
 - Yq{v}^{3}
 + \left( {\frac {51}{16}}
   + \left( -{\frac {7}{16}}+{\frac {15}{16}}\,{Y}^{2} \right)
      {q}^{2} \right) {v}^{4}
\nonumber \\ &&
 - {\frac {43}{4}}\,Yq{v}^{5}
 + \left( {\frac {81}{4}}+ \left( {\frac {153}{8}}\,{Y}^{2}
    - \frac{11}{2} \right) {q}^{2} \right) {v}^{6}
\biggr\}
\nonumber \\ &&
+ e^5 \biggl\{
 {\frac {5}{16}}
 + \frac{{v}^{2}}4
 - \frac{Yq{v}^{3}}{2}
 + \left( {\frac {69}{64}}
     +\left( -{\frac {13}{64}}+{\frac {29}{64}}\,{Y}^{2}
      \right) {q}^{2} \right) {v}^{4}
\nonumber \\ &&
 - {\frac {53}{16}}\,Yq{v}^{5}
 + \left( {\frac {135}{32}}
    + \left( {\frac {43}{8}}\,{Y}^{2}-{\frac {69}{32}}
    \right) {q}^{2} \right) {v}^{6}
\biggr\}, \\
\alpha_4 &=&
e^4 \biggl\{
 \frac{1}{8}
 + \frac{3}{8}{v}^{2}
 - \frac{3}{4}Yq{v}^{3}
 + \left( {\frac {41}{16}}
   + \left( -{\frac {5}{16}}+{\frac {11}{16}}\,{Y}^{2}
       \right) {q}^{2} \right) {v}^{4}
\nonumber \\ &&
 - {\frac {35}{4}}\,Yq{v}^{5}
 + \left( {\frac {277}{16}}
    + \left( {\frac {249}{16}}\,{Y}^{2}-{\frac {69}{16}} \right)
     {q}^{2} \right) {v}^{6}
\biggr\}
\nonumber \\ &&
+ e^6 \biggl\{
 \frac{3}{16}
 + {\frac {5}{16}}\,{v}^{2}
 - \frac{5}{8} Yq{v}^{3}
 + \left( {\frac {27}{16}}
    + \left( -\frac{1}{4}+{\frac {9}{16}}\,{Y}^{2} \right) {q}^{2}
   \right) {v}^{4}
\nonumber \\ &&
 - \frac{11}{2}Yq{v}^{5}
 + \left( 9+ \left( {\frac {75}{8}}\,{Y}^{2}-{\frac {49}{16}}
     \right) {q}^{2} \right) {v}^{6}
\biggr\}, \\
\alpha_5 &=&
e^5 \biggl\{
 \frac{1}{16}
 + \frac{{v}^{2}}{4}
 - \frac{Yq}{2}{v}^{3}
 + \left( {\frac {117}{64}}+ \left( -{\frac {13}{64}}
     +{\frac {29}{64}}\,{Y}^{2} \right) {q}^{2} \right) {v}^{4}
\nonumber \\ &&
 - {\frac {101}{16}}\,Yq{v}^{5}
 +  \left( {\frac {419}{32}}+ \left( {\frac {45}{4}}\,{Y}^{2}
     -{\frac {97}{32}} \right) {q}^{2} \right) {v}^{6}
\biggr\}, \\
\alpha_6 &=&
e^6 \biggl\{
 \frac{1}{32}
 + {\frac {5}{32}}\,{v}^{2}
 - {\frac {5}{16}}\,Yq{v}^{3}
 + \left( {\frac {39}{32}}
     + \left( -\frac{1}{8}+{\frac {9}{32}}\,{Y}^{2}
     \right) {q}^{2} \right) {v}^{4}
\nonumber \\ &&
 - {\frac {17}{4}}\,Yq{v}^{5}
 + \left( {\frac {295}{32}}+ \left( {\frac {243}{32}}\,{Y}^{2}-2
     \right) {q}^{2} \right) {v}^{6}
\biggr\}, \\
\alpha_n &=& O(e^n) \quad {\rm for} \  n \geq 7.
\end{eqnarray}

\subsection{Longitudinal component}
\begin{eqnarray}
\beta_0 &=& 0, \\
\beta_1 &=&
1
+ \left( \frac{1}{16}-{\frac {9}{16}}\,{Y}^{2} \right)
  {q}^{2}{v}^{4}
+ \left( -\frac{1}{4} + \frac{9}{4}{Y}^{2} \right)
  {q}^{2}{v}^{6}
\nonumber \\ &&
+ e^2 \biggl\{
 \left( -\frac{1}{16}+{\frac {9}{16}}\,{Y}^{2} \right)
  {q}^{2}{v}^{4}
 + \left( -\frac{9}{4}{Y}^{2}+\frac{1}{4} \right)
   {q}^{2}{v}^{6}
\biggr\}, \\
\beta_2 &=& 0, \\
\beta_3 &=&
 \frac{1-Y^2}{16} {q}^{2}{v}^{4}
- \frac{1-Y^2}{4} {q}^{2}{v}^{6}
+ e^2 \biggl\{
 - \frac{1-Y^2}{16} {q}^{2}{v}^{4}
 + \frac{1-Y^2}{4} {q}^{2}{v}^{6}
\biggr\}, \\
\beta_n &=& \left \{
\begin{array}{ll}
0 & (n{\rm : even}) \\
O(v^{2n-2}) & (n{\rm : odd})
\end{array}
\right.
\end{eqnarray}

\subsection{$r$-part of the temporal component}
\begin{eqnarray}
\frac{v}{p}\, \tilde{t}_1^{(r)} &=&
e \Bigl\{
 2
 + 4\,{v}^{2}
 - 6\,Yq{v}^{3}
 + \left( 17+ \left( 4\,{Y}^{2}-1 \right) {q}^{2} \right) {v}^{4}
 - 54\,Yq{v}^{5}
\nonumber \\ &&
 + \left( 88+ \left( 84\,{Y}^{2}-20
   \right) {q}^{2} \right) {v}^{6}
\Bigr\}
\nonumber \\ &&
+ e^3 \biggl\{
 3
 + 3\,{v}^{2}
 - 4\,Yq{v}^{3}
 + \left( {\frac {77}{8}}+ \left( {\frac {21}{
     8}}\,{Y}^{2} - \frac{5}{8} \right) {q}^{2} \right) {v}^{4}
 - {\frac {57}{2}}\,Yq{v}^{5}
\nonumber \\ &&
 + \left( {\frac {173}{4}}+ \left( 42\,{Y}^{2}-{\frac {51}{4}}
    \right) {q}^{2} \right) {v}^{6}
\biggr\}
\nonumber \\ &&
+ e^5 \biggl\{
 {\frac {15}{4}}
 + \frac{5}{2} {v}^{2}
 - {\frac {13}{4}}\,Yq{v}^{3}
 + \left( \frac{15}{2}+\left( {\frac {17}{8}}\,{Y}^{2}
     -\frac{1}{2}
     \right) {q}^{2} \right) {v}^{4}
 - {\frac {45}{2}}\,Yq{v}^{5}
\nonumber \\ &&
 + \left( {\frac {67}{2}} + \left( {\frac {133}{4}}\,{Y}^{2}
    -10 \right) {q}^{2} \right) {v}^{6}
\biggr\}, \\
\frac{v}{p}\, \tilde{t}_2^{(r)} &=&
e^2 \biggl\{
 \frac{3}{4}
 + \frac{7}{4} {v}^{2}
 - {\frac {13}{4}}\,Yq{v}^{3}
 + \left( {\frac {81}{8}}+\left( 5/2\,{Y}^{2}
     -{\frac {7}{8}} \right) {q}^{2} \right) {v}^{4}
 - {\frac {135}{4}}\,Yq{v}^{5}
\nonumber \\ &&
 + \left( {\frac {499}{8}}+ \left( 55\,{Y}^{2}-{\frac {113}{8}}
     \right) {q}^{2} \right) {v}^{6}
\biggr\}
\nonumber \\ &&
+ e^4 \biggl\{
 \frac{5}{4}
 + \frac{7}{4} {v}^{2}
 -3\,Yq{v}^{3}
 + \left( {\frac {131}{16}}+ \left( {\frac {37}{16}}\,{Y}^{2}
     -{\frac {13}{16}} \right) {q}^{2} \right) {v}^{4}
 - {\frac {103}{4}}\,Yq{v}^{5}
\nonumber \\ &&
 + \left( {\frac {691}{16}}+ \left( {\frac {655}{16}}\,{Y}^{2}
     -{\frac {197}{16}} \right) {q}^{2} \right) {v}^{6}
\biggr\}
\nonumber \\ &&
+ e^6 \biggl\{
 {\frac {105}{64}}
 + {\frac {105}{64}}\,{v}^{2}
 - {\frac {175}{64}}\,Yq{v}^{3}
 + \left( {\frac {905}{128}}+ \left( {\frac {135}{64}}\,{Y}^{2}
     -{\frac {95}{128}} \right) {q}^{2} \right) {v}^{4}
 - {\frac {1413}{64}}\,Yq{v}^{5}
\nonumber \\ &&
 + \left( {\frac {4591}{128}}+ \left( {\frac {2241}{64}}\,{Y}^{2}
     -{\frac {1389}{128}} \right) {q}^{2} \right) {v}^{6}
\biggr\}, \\
\frac{v}{p}\, \tilde{t}_3^{(r)} &=&
e^3 \biggl\{
 \frac{1}{3}
 + {v}^{2}
 - 2\,Yq{v}^{3}
 + \left( {\frac {53}{8}}+ \left( {\frac {13}{8}}\,{Y}^{2}
     - \frac{5}{8} \right) {q}^{2} \right) {v}^{4}
 - {\frac {45}{2}}\,Yq{v}^{5}
\nonumber \\ &&
 + \left( {\frac {523}{12}}+ \left( 38\,{Y}^{2}-{\frac {39}{4}}
     \right) {q}^{2} \right) {v}^{6}
\biggr\}
\nonumber \\ &&
+ e^5 \biggl\{
 \frac{5}{8}
 + \frac{5}{4} {v}^{2}
 - {\frac {19}{8}}\,Yq{v}^{3}
 + \left( 7+ \left( {\frac {31}{16}}\,{Y}^{2}
     - \frac{3}{4} \right) {q}^{2} \right) {v}^{4}
 - {\frac {91}{4}}\,Yq{v}^{5}
\nonumber \\ &&
 + \left( {\frac {647}{16}}+ \left( {\frac {601}{16}}\,{Y}^{2}
     -{\frac {175}{16}} \right) {q}^{2} \right) {v}^{6}
\biggr\}, \\
\frac{v}{p}\, \tilde{t}_4^{(r)} &=&
e^4 \biggl\{
 {\frac {5}{32}}
 + {\frac {19}{32}}\,{v}^{2}
 - {\frac {39}{32}}\,Yq{v}^{3}
 + \left( {\frac {137}{32}}+ \left( {\frac {65}{64}}\,{Y}^{2}
     -{\frac {13}{32}} \right) {q}^{2} \right) {v}^{4}
 - {\frac {473}{32}}\,Yq{v}^{5}
\nonumber \\ &&
 + \left( {\frac {957}{32}}+ \left( {\frac {1631}{64}}\,{Y}^{2}
     -{\frac {207}{32}} \right) {q}^{2} \right) {v}^{6}
\biggr\}
\nonumber \\ &&
+ e^6 \biggl\{
 {\frac {21}{64}}
 + {\frac {57}{64}}\,{v}^{2}
 - {\frac {113}{64}}\,Yq{v}^{3}
 + \left( {\frac {89}{16}}+ \left(
     {\frac {189}{128}}\,{Y}^{2}-{\frac {19}{32}} \right) {q}^{2}
   \right) {v}^{4}
 - {\frac {1185}{64}}\,Yq{v}^{5}
\nonumber \\ &&
 + \left( {\frac {553}{16}}+ \left( {\frac {4005}{128}}\,{Y}^{2}
     -{\frac {141}{16}} \right) {q}^{2} \right) {v}^{6}
\biggr\}, \\
\frac{v}{p}\, \tilde{t}_5^{(r)} &=&
e^5 \biggl\{
 {\frac {3}{40}}
 + {\frac {7}{20}}\,{v}^{2}
 -{\frac {29}{40}}\,Yq{v}^{3}
 + \left( {\frac {27}{10}}+ \left( {\frac {49}{80}}\,{Y}^{2}
     -1/4 \right) {q}^{2} \right) {v}^{4}
 - {\frac {189}{20}}\,Yq{v}^{5}
\nonumber \\ &&
 + \left( {\frac {319}{16}}+ \left( {\frac {1321}{80}}
     {Y}^{2}-{\frac {331}{80}} \right) {q}^{2}
   \right) {v}^{6}
\biggr\}, \\
\frac{v}{p}\, \tilde{t}_6^{(r)} &=&
e^6 \biggl\{
 {\frac {7}{192}}
 + {\frac {13}{64}}\,{v}^{2}
 - {\frac {27}{64}}\,Yq{v}^{3}
 + \left( {\frac {213}{128}}+ \left(
     {\frac {23}{64}}\,{Y}^{2}-{\frac {19}{128}}
     \right) {q}^{2} \right) {v}^{4}
 - {\frac {377}{64}}\,Yq{v}^{5}
\nonumber \\ &&
 + \left( {\frac {4969}{384}}+ \left(
     {\frac {665}{64}}\,{Y}^{2}-{\frac {329}{128}}
     \right) {q}^{2} \right) {v}^{6}
\biggr\}, \\
\frac{v}{p}\, \tilde{t}_n^{(r)} &=&
O(e^n) \quad ({\rm for} \ n \ge 7).
\end{eqnarray}

\subsection{$\theta$-part of the temporal component}
\begin{eqnarray}
\frac{1}{p}\, \tilde{t}_1^{(\theta)} &=& 0, \\
\frac{1}{p}\, \tilde{t}_2^{(\theta)} &=&
 \frac{(Y^2-1)}{4} q^2 v^3
 - \frac{(Y^2-1)}{2} q^2 v^5
 + \frac{Y(Y^2-1)}{4} (3+e^2) q^3 v^6, \\
\frac{1}{p}\, \tilde{t}_n^{(\theta)} &=& \left \{
\begin{array}{ll}
0 & (n{\rm : odd}) \\
O(v^{2n-1}) & (n{\rm : even})
\end{array}
\right.
\end{eqnarray}

\subsection{$r$-part of the azimuthal component}
\begin{eqnarray}
\tilde\varphi_1^{(r)} &=&
e \left\{
 - 2\,q{v}^{3}
 + 2\,Y{q}^{2}{v}^{4}
 - 10\,q{v}^{5}
 + 18\,Y{q}^{2}{v}^{6}
\right\}, \\
\tilde\varphi_2^{(r)} &=&
e^2 \left\{
 - \frac{1}{4}\,Y{q}^{2}{v}^{4}
 + \frac{1}{2}\,q{v}^{5}
 - \frac{3}{4}\,Y{q}^{2}{v}^{6}
\right\}, \\
\tilde\varphi_{n}^{(r)} &=& \left \{
\begin{array}{ll}
O(v^{2n+1}) & (n{\rm : odd}) \\
O(v^{2n}) & (n{\rm : even})
\end{array}
\right.
\end{eqnarray}

\subsection{$\theta$-part of the azimuthal component}
\begin{eqnarray}
X_0^\Re &=&
\left\{
 \frac{1+Y}{2}
 - \frac{\left( 9\,Y-1 \right)  \left( {Y}^{2}-1 \right)}{32}
    {q}^{2}{v}^{4}
 + \frac{\left( 9\,Y-1 \right)  \left( {Y}^{2}-1 \right)}{8}
   {q}^{2}{v}^{6}
\right\}
\nonumber \\ &&
+ e^2 \left\{
 \frac{\left( 9\,Y-1 \right)  \left( {Y}^{2}-1 \right)}{32}
 {q}^{2}{v}^{4}
 - \frac{\left( 9\,Y-1 \right)  \left( {Y}^{2}-1 \right)}{8}
  {q}^{2}{v}^{6}
\right\}, \\
X_1^\Re &=& 0, \\
X_2^\Re &=&
\left\{
 \frac{1-Y}{2}
 + \frac{Y \left( {Y}^{2}-1 \right)}{4} {q}^{2}{v}^{4}
 - Y \left( {Y}^{2}-1 \right) {q}^{2}{v}^{6}
\right\}
\nonumber \\ &&
+ e^2 \left\{
 - \frac{Y \left( {Y}^{2}-1 \right)}{4} {q}^{2}{v}^{4}
 +Y \left( {Y}^{2}-1 \right) {q}^{2}{v}^{6}
\right\}, \\
X_3^\Re &=& 0, \\
X_4^\Re &=&
\left\{
 \frac{\left( Y+1 \right)  \left( Y-1 \right)^{2}}{32}
 {q}^{2}{v}^{4}
 - \frac{\left( Y+1 \right)  \left( Y-1 \right)^{2}}{8}
 {q}^{2}{v}^{6}
\right\}
\nonumber \\ &&
+ e^2 \left\{
 - \frac{\left( Y+1 \right)  \left( Y-1 \right)^{2}}{32}
   {q}^{2}{v}^{4}
 + \frac{\left( Y+1 \right)  \left( Y-1 \right)^{2}}{8}
  {q}^{2}{v}^{6}
\right\}, \\
X_n^\Re &=& \left \{
\begin{array}{ll}
0 & (n{\rm : odd}) \\
O(v^{2n-4}) & (n \ge 6{\rm : even})
\end{array}
\right., \\
X_0^\Im &=& 0, \\
X_1^\Im &=& 0, \\
X_2^\Im &=&
\left\{
 \frac{Y-1}{2}
 - \frac{\left( 5\,Y+1 \right)  \left( {Y}^{2}-1 \right)}{16}
  {q}^{2}{v}^{4}
 + \frac{\left( 5\,Y+1 \right)  \left( {Y}^{2}-1 \right)}{4}
  {q}^{2}{v}^{6}
\right\}
\nonumber \\ &&
+ e^2 \left\{
 \frac{\left( 5\,Y+1 \right)  \left( {Y}^{2}-1 \right)}{16}
 {q}^{2}{v}^{4}
 - \frac{\left( 5\,Y+1 \right)  \left( {Y}^{2}-1 \right)}{4}
  {q}^{2}{v}^{6}
\right\}, \\
X_3^\Im &=& 0, \\
X_4^\Im &=&
\left\{
 - \frac{\left( Y+1 \right)  \left( Y-1 \right)^{2}}{32}
   {q}^{2}{v}^{4}
 + \frac{\left( Y+1 \right)  \left( Y-1 \right)^{2}}{8}
   {q}^{2}{v}^{6}
\right\}
\nonumber \\ &&
+ e^2 \left\{
 \frac{\left( Y+1 \right)  \left( Y-1 \right)^{2}}{32}
 {q}^{2}{v}^{4}
 - \frac{\left( Y+1 \right)  \left( Y-1 \right)^{2}}{8}
   {q}^{2}{v}^{6}
\right\}, \\
X_n^\Im &=& \left \{
\begin{array}{ll}
0 & (n{\rm : odd}) \\
O(v^{2n-4}) & (n \ge 6{\rm : even})
\end{array}
\right..
\end{eqnarray}

\section{Secular evolution of the orbital parameters,
$v$, $e$, and $Y$} \label{App:orbit_evolv}
An alternative set of the orbital parameters, $J=\{v,e,Y\}$, is also
useful to specify the orbit. The secular changes of the parameters
can be derived from those of $I=\{E,L,C\}$, as 
\begin{equation}
\left<\frac{dJ}{dt}\right>_t =
\sum_{I=E,L,C} (G^{-1})^J_I\,\left<\frac{dI}{dt}\right>_t,
\label{eq:trans_params}
\end{equation}
where $G^I_J={\partial (E,L,C)}/{\partial (v,e,Y)}$ is the Jacobian
matrix for the transformation from $\{E,L,C\}$ to $\{v,e,Y\}$
\footnote{It should be noted that, to calculate the Jacobian matrix
up to $O(e^6)$, one need to calculate $\{E,L,C\}$ up to $O(e^8)$
since the leading terms do not depend on $e$ and then the relative
orders of accuracy of their first derivatives with the
eccentricity of the Jacobian matrix in Eq.~(\ref{eq:trans_params})
are reduced by $O(e^2)$.
For a similar reason, one also need to calculate $E$ up to 5PN order
because the relative PN order of $\partial E/\partial v$ is reduced by
$O(v^2)$ compared to $E$.}.

Substituting the 3PN $O(e^6)$ formulae of $\langle dI/dt \rangle_t^\infty$
shown in Sec.~\ref{sec:result} into the above relation,
we obtain the secular changes of $\{v,e,Y\}$
associated with the flux of gravitational waves to infinity as
\begin{eqnarray}
 \left<\frac{dv}{dt}\right>^\infty_t &=&
 \left(\frac{dv}{dt}\right)_{\rm N}
\biggl[
1+\frac{7}{8}{e}^2
+\left\{-\frac{743}{336}-\frac{55}{21}{e}^2+{\frac {8539}{2688}}\,{e}^{4}\right\}v^2
\nonumber \\ && 
+ \biggl\{ 4\,\pi-{\frac {133}{12}}\,Yq + \left( {\frac {97}{8}}\,\pi 
-{\frac {379}{24}}\,Yq \right) {e}^{2}
\nonumber \\ && \hspace{0.5cm}
+ 
\left( {\frac {49}{32}}\,\pi -{\frac {475}{96}}\,Yq \right) {e}^{4}
-{\frac {49}{4608}}\,\pi \,{e}^{6} \biggr\} {v}^{3}
\nonumber \\  && 
+ \biggl\{ {\frac {34103}{18144}}-{\frac {329}{96}}\,{q}^{2}
+{\frac {815}{96}}\,{Y}^{2}{q}^{2}
+ \left( -{\frac {526955}{12096}}-{\frac {
929}{96}}\,{q}^{2}+{\frac {477}{32}}\,{Y}^{2}{q}^{2} \right) {e}^{2}
\nonumber \\ && \hspace{0.5cm}
+ \left( -{\frac {1232809}{48384}}-{\frac {1051}{768}}\,{q}^{2}+{\frac 
{999}{256}}\,{Y}^{2}{q}^{2} \right) {e}^{4}+{\frac {105925}{16128}}\,{
e}^{6} \biggr\} {v}^{4}
\nonumber \\ && 
+ \biggl\{ -{\frac {4159}{672}}\,\pi -{\frac {1451}{56}}\,Yq
+ \left( -{\frac {48809}{1344}}\,\pi -{\frac {1043}{96}}\,Yq \right) {e}^{2}
\nonumber \\ && \hspace{0.5cm}
+ \left( {\frac {679957}{43008}}\,\pi -{\frac {15623}{336}}\,Yq \right) {e}^{4}
+ \left( {\frac {4005097}{774144}}\,\pi -{
\frac {35569}{1792}}\,Yq \right) {e}^{6} \biggr\} {v}^{5}
\nonumber \\ && 
+ \biggl\{ 
 {\frac {16447322263}{139708800}}
+\frac{16}{3}\,{\pi }^{2}
-{\frac {1712}{105}}\,\gamma-{\frac {3424}{105}}\,\ln  \left( 2 \right) 
-{\frac {331}{192}}\,{q}^{2}
\nonumber \\ && \hspace{0.5cm}
-{\frac {289}{6}}\,\pi \,Yq+{\frac {145759}{1344}}\,{Y}^{2}{q}^{2}
\nonumber \\ && \hspace{0.5cm}
+ \biggl( {\frac {8901670423}{11642400}}+{\frac {229}{6}}\,{\pi }^
{2}-{\frac {24503}{210}}\,\gamma+{\frac {1391}{30}}\,\ln  \left( 2
 \right) -{\frac {78003}{280}}\,\ln  \left( 3 \right) 
\nonumber \\  && \hspace{1.0cm}
+{\frac {2129}{42}}\,{q}^{2}-{\frac {4225}{24}}\,\pi \,Yq
+{\frac {27191}{224}}\,{Y}^{2}{q}^{2} \biggr) {e}^{2}
\nonumber \\ && \hspace{0.5cm}
+ \biggl( {\frac {269418340489}{372556800}}+{
\frac {109}{4}}\,{\pi }^{2}+{\frac {3042117}{1120}}\,\ln  \left( 3
 \right) -{\frac {11663}{140}}\,\gamma
\nonumber \\ && \hspace{1.0cm}
-{\frac {418049}{84}}\,\ln  \left( 2 \right) 
-{\frac {56239}{10752}}\,{q}^{2}-{\frac {17113}{192}
}\,\pi \,Yq+{\frac {414439}{3584}}\,{Y}^{2}{q}^{2} \biggr) {e}^{4}
\nonumber \\ && \hspace{0.5cm}
+ \biggl( {\frac {174289281}{862400}}-{\frac {1044921875}{96768}}\,\ln 
 \left( 5 \right)
 +{\frac {23}{16}}\,{\pi }^{2}
-{\frac {42667641}{3584}}\,\ln  \left( 3 \right) 
\nonumber \\ && \hspace{1.0cm}
+{\frac {94138279}{2160}}\,\ln  \left( 2 \right) 
-{\frac {2461}{560}}\,\gamma-{\frac {3571}{3584}}\,{q}^{2}-{
\frac {108577}{13824}}\,\pi \,Yq
\nonumber \\ && \hspace{1.0cm}
+{\frac {41071}{1536}}\,{Y}^{2}{q}^{2}
 \biggr) {e}^{6}
\nonumber \\ && \hspace{0.5cm}
- \left( {\frac {1712}{105}}+{\frac {24503}{210}}\,{e}^{2}
+{\frac {11663}{140}}\,{e}^{4}+{\frac {2461}{560}}\,{e}^{6}
 \right) \ln v \biggr\} {v}^{6}
\biggr],
\label{eq:dotv8} \\
\left< \frac{d{e}}{dt}\right>^\infty_t  &=&
 \left(\frac{de}{dt}\right)_{\rm N}
\biggl[
1+{\frac {121}{304}}\,{e}^{2}+ 
\left\{ -{\frac {6849}{2128}}-{\frac {
2325}{2128}}\,{e}^{2}+{\frac {22579}{17024}}\,{e}^{4} \right\} {v}^{2}
\nonumber \\ && 
+ \biggl\{ {\frac {985}{152}}\,\pi -{\frac {879}{76}}\,Yq
+ \left( {
\frac {5969}{608}}\,\pi -{\frac {699}{76}}\,Yq \right) {e}^{2}
\nonumber \\ && \hspace{0.5cm}
+ \left( {\frac {24217}{29184}}\,\pi -{\frac {1313}{608}}\,Yq \right) {e}^{4}
\biggr\} {v}^{3}
\nonumber \\ && 
+ \biggl\{ -{
\frac {286397}{38304}}-{\frac {3179}{608}}\,{q}^{2}+{\frac {5869}{608}
}\,{Y}^{2}{q}^{2}
\nonumber \\ && \hspace{0.5cm}
+ \left( -{\frac {2070667}{51072}}-{\frac {8925}{1216
}}\,{q}^{2}+{\frac {633}{64}}\,{Y}^{2}{q}^{2} \right) {e}^{2}
\nonumber \\ && \hspace{0.5cm}
+ \left( 
-{\frac {3506201}{306432}}-{\frac {3191}{4864}}\,{q}^{2}+{\frac {9009}
{4864}}\,{Y}^{2}{q}^{2} \right) {e}^{4}
\biggr\} {v}^{4}
\nonumber \\ && 
+ \biggl\{ -{\frac {1903}{304}}\,Yq-{\frac {87947}{
4256}}\,\pi + \left( -{\frac {3539537}{68096}}\,\pi -{\frac {93931}{
8512}}\,Yq \right) {e}^{2}
\nonumber \\ && \hspace{0.5cm}
+ \left( {\frac {5678971}{817152}}\,\pi -{
\frac {442811}{17024}}\,Yq \right) {e}^{4}
\biggr\} {v}^{5}
\nonumber \\ && 
+ \biggl\{ -{\frac {82283}{1995}}
\,\gamma-{\frac {11021}{285}}\,\ln  \left( 2 \right) -{\frac {234009}{
5320}}\,\ln  \left( 3 \right) +{\frac {11224646611}{46569600}}+{\frac 
{769}{57}}\,{\pi }^{2}
\nonumber \\ && \hspace{0.5cm}
+{\frac {180255}{8512}}\,{q}^{2}-{\frac {11809}{
152}}\,\pi \,Yq+{\frac {598987}{8512}}\,{Y}^{2}{q}^{2}
\nonumber \\ && \hspace{0.5cm}
+ \biggl( {\frac 
{927800711807}{884822400}}-{\frac {2982946}{1995}}\,\ln  \left( 2
 \right) +{\frac {2782}{57}}\,{\pi }^{2}+{\frac {1638063}{3040}}\,\ln 
 \left( 3 \right) 
\nonumber \\ && \hspace{1.0cm}
-{\frac {297674}{1995}}\,\gamma
+{\frac {536653}{8512
}}\,{q}^{2}-{\frac {91375}{608}}\,\pi \,Yq+{\frac {356845}{8512}}\,{Y}
^{2}{q}^{2} \biggr) {e}^{2}
\nonumber \\ && \hspace{0.5cm}
+ \biggl( {\frac {190310746553}{262169600}}-
{\frac {1147147}{15960}}\,\gamma+{\frac {10721}{456}}\,{\pi }^{2}-{
\frac {1022385321}{340480}}\,\ln  \left( 3 \right) 
\nonumber \\ && \hspace{1.0cm}
+{\frac {760314287}{47880}}\,\ln  \left( 2 \right) 
-{\frac {1044921875}{204288}}\,\ln 
 \left( 5 \right) +{\frac {56509}{9728}}\,{q}^{2}
\nonumber \\ && \hspace{1.0cm}
-{\frac {1739605}{
29184}}\,\pi \,Yq+{\frac {3248951}{68096}}\,{Y}^{2}{q}^{2} \biggr) {e}^{4}
\nonumber \\ && \hspace{0.5cm}
- \left( {\frac {82283}{1995}}+{\frac {297674}{1995}}\,{e}^{2}
+{\frac {1147147}{15960}}\,{e}^{4}
\right) \ln v \biggr\} {v}^{6}
\biggr],\label{eq:dote8}\\
\left< \frac{dY}{dt}\right>^\infty_t &=&
 \left(\frac{dY}{dt}\right)_{\rm N}
\biggl[
1+{\frac {189}{61}}\,{e}^{2}+{\frac {285}{488}}\,{e}^{4}
\nonumber \\ && 
+\biggl\{ -{\frac {13}{244}}\,Yq-{\frac {277}{244}}\,Yq
{e}^{2}-{\frac {1055}{1952}}\,Yq{e}^{4} \biggr\} {v}
\nonumber \\ && 
+ \left\{ -{
\frac {10461}{1708}}-{\frac {83723}{3416}}\,{e}^{2}-{\frac {21261}{
13664}}\,{e}^{4}+{\frac {49503}{27328}}\,{e}^{6} \right\} {v}^{2}
\nonumber \\ && 
+ \biggl\{ {\frac {290}{61}}\,\pi -{\frac {12755}{3416}}\,Yq+ \left( {
\frac {1990}{61}}\,\pi -{\frac {27331}{1708}}\,Yq \right) {e}^{2}
\nonumber \\ && \hspace{0.5cm}
+ \left( {\frac {21947}{976}}\,\pi -{\frac {540161}{27328}}\,Yq
 \right) {e}^{4}+ \left( {\frac {38747}{35136}}\,\pi -{\frac {140001}{
27328}}\,Yq \right) {e}^{6} \biggr\} {v}^{3}
\biggr],\label{eq:dotY8}
\end{eqnarray}
where the leading contributions are given by
\begin{eqnarray}
\left( \frac{dv}{dt} \right)_{\rm N} &=&
\frac{32}{5}\left(\frac{\mu}{M^2}\right)v^9(1-{e}^2)^{3/2},
\nonumber \\
\left( \frac{de}{dt} \right)_{\rm N} &=&
-\frac{304}{15}\left(\frac{\mu}{M^2}\right)v^8{e}\,(1-{e}^2)^{3/2},
\nonumber \\
\left( \frac{dY}{dt} \right)_{\rm N} &=&
 -\frac{244}{15}\left(\frac{\mu}{M^2}\right)v^{11} q\,(1-{e}^2)^{3/2}(1-Y^2).
\label{eq:dotIN}
\end{eqnarray}

In the same way, substituting the 3.5PN $O(e^6)$ formulae of
$\langle dI/dt \rangle_t^{{\rm H}}$ shown in Sec.~\ref{sec:result}
into Eq.~(\ref{eq:trans_params}), we obtain
the secular changes of $\{v,e,Y\}$ associated with the flux of gravitational
waves to the horizon as
\begin{eqnarray}
 \left<\frac{dv}{dt}\right>^{\rm H}_t &=&
 \left(\frac{dv}{dt}\right)_{\rm N}
\biggl[
-{\frac {1}{256}} \left\{ 8+24\,{e}^{2}+3\,{e}^{4} \right\}  
\left\{8+9\,{q}^{2}+15\,{Y}^{2}{q}^{2} \right\} \,Y\,q\,v^5
\nonumber \\ && 
\biggl\{
-{\frac {11}{8}}-{\frac {189}{64}}\,{q}^{2}-{\frac {15}{64}}\,{Y}^{2}{q}^{2}
+ \left( -{\frac {69}{8}}-{\frac {81}{4}}\,{q}^{2}
+{\frac {45}{32}}\,{Y}^{2}{q}^{2} \right) {e}^{2}
\nonumber \\ && \hspace{0.5cm}
+ \left( -{\frac {381}{64}}-{\frac {7479}{512}}\,{q}^{2}
+{\frac {1035}{512}}\,{Y}^{2}{q}^{2} \right) {e}^{4}
\nonumber \\ && \hspace{0.5cm}
+ 
\left( -{\frac {11}{32}}-{\frac {423}{512}}\,{q}^{2}
+{\frac {45}{512}}\,{Y}^{2}{q}^{2} \right) {e}^{6}
\biggr\} \,Y\,q\,v^7
\biggr],\label{eq:dotvH}\\
 \left< \frac{d{e}}{dt}\right>^{\rm H}_t  &=&
 \left(\frac{de}{dt}\right)_{\rm N}
\biggl[
-{\frac {33}{4864}} \left\{ 8+12\,{e}^{2}+{e}^{4} \right\}  
\left\{ 8+9\,{q}^{2}+15\,{Y}^{2}{q}^{2} \right\} \,Yq\,v^5
\nonumber \\ && 
\biggl\{
-{\frac {453}{152}}-{\frac {8127}{1216}}\,{q}^{2}-{\frac {45}{
1216}}\,{Y}^{2}{q}^{2}+ \left( -{\frac {2979}{304}}-{\frac {28593}
{1216}}\,{q}^{2}+{\frac {1485}{608}}\,{Y}^{2}{q}^{2} \right) {e}^{2}
\nonumber \\ && \hspace{0.5cm}
+ \left( -{\frac {5649}{1216}}-{\frac {111591}{9728}}\,{q}^{2}+{
\frac {16515}{9728}}\,{Y}^{2}{q}^{2} \right) {e}^{4}
\biggr\}\,Yq\,v^7
\biggr],\label{eq:doteH}\\
\left< \frac{dY}{dt}\right>^{\rm H}_t &=&
 \left(\frac{dY}{dt}\right)_{\rm N}
\biggl[
-{\frac {3}{7808}}\, \left\{ 8+24\,{e}^{2}+3\,{e}^{4} \right\}  
\left\{ 16+33\,{q}^{2}+15\,{Y}^{2}{q}^{2} \right\} \,v^2
\nonumber \\ && 
\biggl\{
-{\frac {51}{122}}
+{\frac {585}{1952}}\,{Y}^{2}{q}^{2}-{\frac {1953}{1952}}\,{q}^{2}
+\left( -{\frac {225}{61}}-{\frac {16875}{1952}}\,{q}
^{2}+{\frac {3375}{1952}}\,{Y}^{2}{q}^{2} \right) {e}^{2}
\nonumber \\ && \hspace{0.5cm}
+ \left( -{\frac {2961}{976}}-{\frac {109863}{15616}}\,{q}^{2}+{\frac {16335}{
15616}}\,{Y}^{2}{q}^{2} \right) {e}^{4}
\nonumber \\ && \hspace{0.5cm}
+ \left( -{\frac {171}{976}}-{
\frac {3159}{7808}}\,{q}^{2}+{\frac {405}{7808}}\,{Y}^{2}{q}^{2}
 \right) {e}^{6}
\biggr\}\,v^4
\biggr].\label{eq:dotYH}
\end{eqnarray}
Actually, we can obtain the higher PN results by using the 4PN $O(e^6)$
formulae for the secular changes of $\{E,L,C\}$,
although we do not present them in the text.
The full expressions of $\langle dJ/dt \rangle_t^\infty$ and
$\langle dJ/dt \rangle_t^{{\rm H}}$ for $J=\{v,e,Y\}$
will be available online~\cite{BHPC}.

Here we make a comment on the reliable order of the expansion with respect
to $e$ in $\langle de/dt \rangle_t$. By using Eq.~(\ref{eq:trans_params}),
$\langle de/dt \rangle_t$ can be calculated from the linear combination of
the secular changes of $\{E,L,C\}$. Since the leading order of the
$(e,I)$-component of the inverse Jacobian matrix is $O(1/e)$,
each term in the linear combination is apparently $O(1/e)$.
However, the $O(1/e)$ contribution turns out to vanish due to a
cancellation in taking the combination, and hence $\langle de/dt \rangle_t$
is $O(e)$, which corresponds to the well-known fact that
circular orbits remain circular~\cite{Peters:1964rf,Tagoshi:1995sh}, 
{\it i.e.} $\langle de/dt \rangle_t=0$ when $e=0$.
This cancellation reduces the reliable order in $\langle de/dt \rangle_t$
by $O(e^2)$, compared to the order of $\langle dI/dt \rangle_t$ for
$I=\{E,L,C\}$.
Since we calculate $\langle dI/dt \rangle_t$ up to $O(e^6)$ in this paper,
we can obtain $\langle de/dt \rangle_t$ correctly up to $O(e^4)$ from the
leading order.

\begin{figure}
\includegraphics[width=51mm]{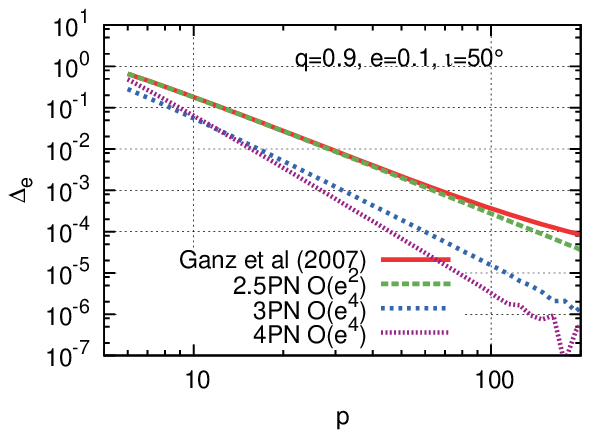}%
\includegraphics[width=51mm]{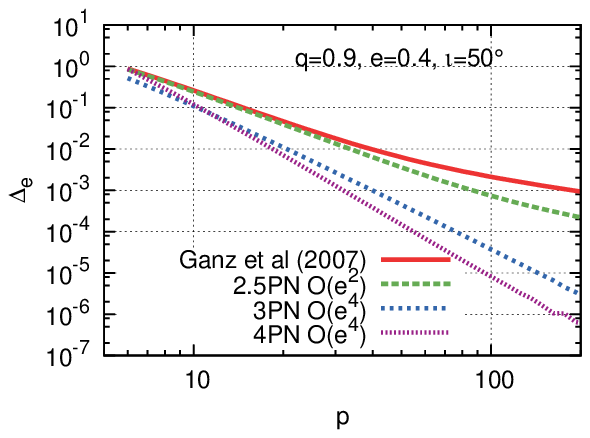}%
\includegraphics[width=51mm]{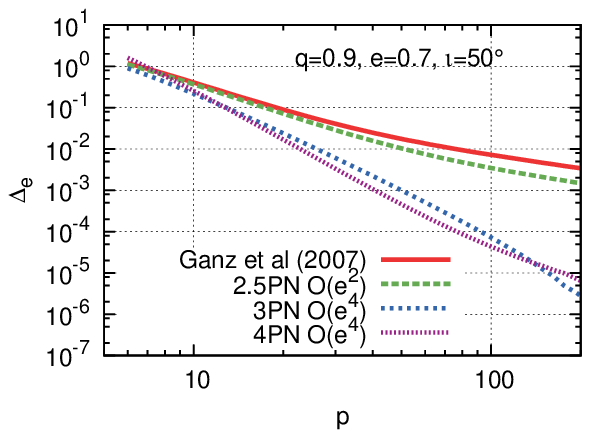}
\caption{The relative errors in the analytic formulae for the secular
change of the orbital eccentricity due to the gravitational waves to
infinity. We plot $\Delta_e$, defined in a similar manner to
Eq.~(\ref{eq:relative_error}), as a function of 
the semi-latus rectum $p$ for $q=0.9$, $e=0.1, 0.4$ and $0.7$ 
(from left to right) and $\iota=50^{\circ}$. 
We truncated the plots at $p=6$ because the relative errors get too large
in $p<6$ to be meaningful.
The relative error in the previous 2.5PN $O(e^2)$
formula given in \cite{Ganz:2007rf} strays off the $p^{-3}$ line
earlier than the 2.5PN $O(e^2)$ formula in this paper. This trend
is clearer for larger $e$.
The relative errors in the 3PN $O(e^4)$ and 4PN $O(e^4)$ formulae fall
off faster than $p^{-3}$ and $p^{-4}$ for small $e$ cases
as expected, while this is not the case for $e=0.7$ because of the
higher order correction of $e$ than $O(e^4)$.
} \label{fig:dedt8_q0.9_e0.1_0.7_inc50}
\end{figure}

$\langle dv/dt \rangle^\infty_t$ and $\langle dY/dt \rangle^\infty_t$
in Eqs.~(\ref{eq:dotv8}) and (\ref{eq:dotY8}) are consistent up to 
the 2.5PN $O(e^2)$ order with the previous results in Ref.~\cite{Ganz:2007rf},
while we find inconsistency in the $O(e^2)$ terms
of the formula for $\langle de/dt \rangle^\infty_t$ in \cite{Ganz:2007rf}.
This may be explained by the reduction in the reliable order mentioned
above: the calculations of $\langle dI/dt \rangle_t$ in \cite{Ganz:2007rf}
are done up to $O(e^2)$, and therefore the resultant formula of
$\langle de/dt \rangle_t$ is reliable only at the leading order.
We can also confirm it numerically. 
In Fig.~\ref{fig:dedt8_q0.9_e0.1_0.7_inc50} we show the relative errors
in the two analytic formulae by comparing to numerical
results~\cite{Fujita:2009us} in a similar manner to
Eq.~(\ref{eq:relative_error}).
It can be found that the relative error in the previous 2.5PN
$O(e^2)$ formula strays out of the expected power law, $p^{-3}$,
earlier than that in our 2.5PN $O(e^2)$ formula.
This trend is clearer for larger eccentricity.
We can also confirm the validity of our formula by seeing that the
relative errors in our 4PN $O(e^4)$ formula falls off faster than
$p^{-4}$.
(The leading PN order of the difference in the $O(e^2)$ terms between
the previous and our formulae is $\frac{39}{38}e^2v^2$. If our
formula contains any error in the $e^2 v^2$ term, the relative
error will not fall off faster than $p^{-4}$ for large $p$.)

A similar reduction in the PN order occurs in the calculation of
$\langle dY/dt \rangle_t^\infty$: although each term in the linear combination
of Eq.~(\ref{eq:trans_params}) for $J=Y$ is $O(v^8)$, the terms at the first
two orders, $O(v^8)$ and $O(v^{10})$, vanish due to a cancellation in taking the
combination. As a result, the leading order of $\langle dY/dt \rangle_t$ is
$O(v^{11})$ and hence the reliable order relative to the leading term is reduced
to $O(v^5)$ (2.5PN order) when we have $\langle dI/dt \rangle_t$ for $I=\{E,L,C\}$
up to $O(v^8)$ (4PN order).

\begin{figure}
\includegraphics[width=51mm]{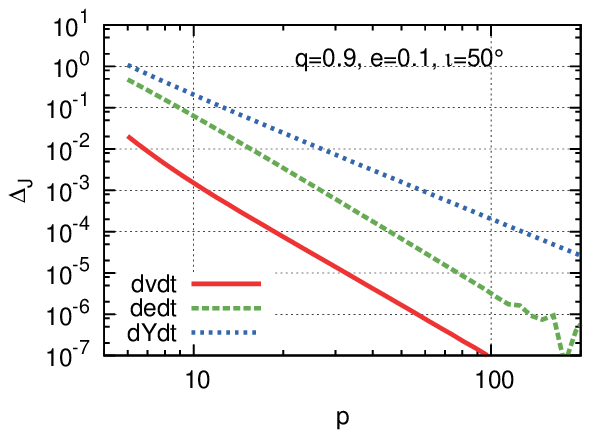}%
\includegraphics[width=51mm]{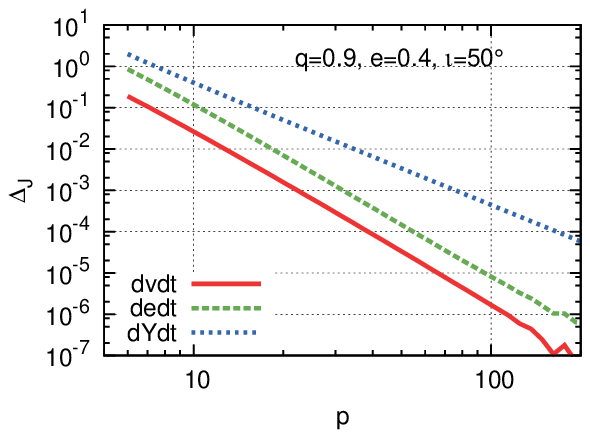}%
\includegraphics[width=51mm]{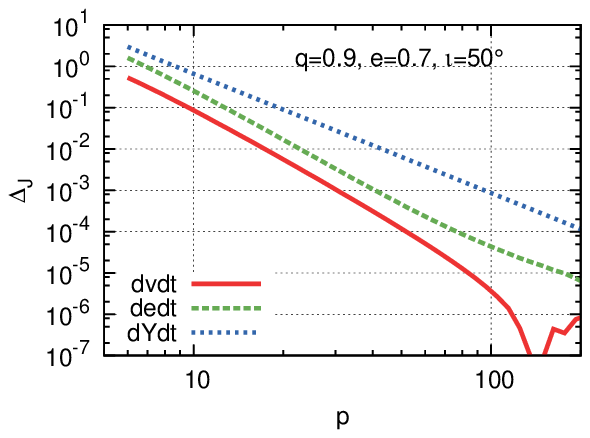}
\caption{The relative errors in the 4PN formulae for the
secular changes of the orbital parameters, $\{v,e,Y\}$.
We plot the relative errors, $\Delta_J$ defined in a similar manner
to Eq.~(\ref{eq:relative_error}), as functions of 
the semi-latus rectum $p$ for $q=0.9$, $e=0.1, 0.4$ and $0.7$ 
(from left to right) and $\iota=50^{\circ}$.
We truncated the plots at $p=6$ because the relative errors get too large
in $p<6$ to be meaningful.
$\Delta_v$ and $\Delta_e$ fall off faster than $p^{-4}$, while
$\Delta_Y$ approximately fall off as $p^{-3}$, slower than $O(p^{-4})$.
This confirms that the relative order of the PN correction of the analytic
formula for $\langle dY/dt \rangle_t$ is reduced from 4PN to 2.5PN
because of the cancellation of the low PN terms.
} \label{fig:dotI8H_q0.9_e0.1_0.4_0.7_inc50}
\end{figure}

In Fig.~\ref{fig:dotI8H_q0.9_e0.1_0.4_0.7_inc50}, we show 
the relative errors in the analytic PN formulae for
the secular changes of the orbital parameters, 
$\{v, e, Y\}$, derived from the 4PN $O(e^6)$ formulae of
$\langle dI/dt \rangle_t$ for $I=\{E,L,C\}$.
Similarly in Fig.~\ref{fig:dotE8H_q0.9_e0.1_0.7_inc20_80}, 
the relative errors in the analytic formulae for $\langle dv/dt \rangle_t$ and
$\langle de/dt \rangle_t$ as functions of the semi-latus rectum $p$ 
fall off faster than $p^{-4}$ when the eccentricity is small.
Observe, however, that the relative error in the analytic formula 
for $\langle dY/dt \rangle_t$ falls off faster than $p^{-5/2}$, but
slower than $p^{-4}$.

From the leading order expressions in Eq.~(\ref{eq:dotIN}), 
one will find the well known fact that
equatorial orbits stay in the equatorial 
plane~\cite{Shibata:1994jx,Sago:2005fn,Ganz:2007rf}, 
{\it i.e.} $\langle dY/dt \rangle_t=0$ when $Y=1$.
In the Schwarzschild case ($q=0$), the secular changes of $v$ and $e$
do not depend on $Y$ in addition to $\langle dY/dt \rangle_t=0$.
This implies that the orbital plane can be fixed on the equatorial
plane ($\theta=\pi/2$) due to the spherical symmetry of
Schwarzschild spacetime.

One will also find that the radiation reaction reduces 
the orbital eccentricity and increases the orbital velocity 
since $(de/dt)_{\rm N}\le 0$ 
and $(dv/dt)_{\rm N}\ge 0$~\cite{Peters:1964rf,Tagoshi:1995sh}, 
while the radiation reaction increases (decreases) the inclination angle 
since $(dY/dt)_{\rm N}\le 0$
($(dY/dt)_{\rm N}\ge 0$) 
when $q\ge 0$ ($q\le 0$)~\cite{Shibata:1994jx,Sago:2005fn,Ganz:2007rf}. 
Moreover, the secular change of the inclination angle is smaller than 
those of the other orbital parameters since 
$\langle d\ln e/d\ln v\rangle_t=O(v^0)$ and 
$\langle d\ln Y/d\ln v\rangle_t=O(v^3)$.


%


\end{document}